\documentclass[10pt,conference,letterpaper]{IEEEtran}

\usepackage[utf8]{inputenc}
\usepackage{times}

\usepackage[T1]{fontenc}
\usepackage[american]{babel}

\usepackage{amsmath}
\usepackage{amssymb}
\usepackage{wasysym}

\usepackage{bm}

\usepackage[frozencache]{minted}
\usepackage{graphicx}
\usepackage[noend]{algorithmic}
\algsetup{indent=1.5em}
\graphicspath{{figures}}

\usepackage{tikz}
\usetikzlibrary{decorations.pathreplacing}
\usetikzlibrary{fit}
\usetikzlibrary{patterns}

\usepackage{pgfplotstable}
\usepackage{pgfplots}
\usepgfplotslibrary{units}
\pgfplotsset{compat=1.14}

\usepackage{fancyvrb}
\usepackage{csquotes} 
\usepackage{bold-extra}
\usepackage{subcaption}
\usepackage{booktabs}
\usepackage{dblfloatfix}
\usepackage{flushend}
\usepackage[binary-units]{siunitx}

\DeclareSIUnit{\nothing}{\relax}
\DeclareSIUnit{\x}{\times}
\newcommand{\factor}[1]{\SI{#1}{\x}}
\newcommand{\factorrange}[2]{\SIrange[range-phrase={ and }]{#1}{#2}{\x}}

\newcommand{\lbcell}[2][c]{%
  \begin{tabular}[#1]{@{}c@{}}#2\end{tabular}}

\usepackage[inline]{enumitem}
\setlist[itemize]{noitemsep,topsep=0pt,leftmargin=1.5em}

\usepackage[disable]{todonotes}
\newcommand\itodo[1]{\todo[inline]{#1}}

\usepackage{newfloat}
\DeclareFloatingEnvironment[fileext=alg,name=Algorithm]{algorithm}

\PassOptionsToPackage{sortcites,firstinits=true}{biblatex}
\usepackage[%
  backend=biber
]{biblatex}
\definecolor{ETHa}{RGB}{31,64,122}      
\definecolor{ETHb}{RGB}{72,90,44}       
\definecolor{ETHc}{RGB}{18,105,176}     
\definecolor{ETHd}{RGB}{114,121,28}     
\definecolor{ETHe}{RGB}{145,5,106}      
\definecolor{ETHf}{RGB}{111,111,100}    
\definecolor{ETHg}{RGB}{168,50,45}      
\definecolor{ETHh}{RGB}{0,122,150}      
\definecolor{ETHi}{RGB}{149,96,19}      

\definecolor{LinkRed}{cmyk}{0,1,1,0.5}
\PassOptionsToPackage{pdfpagelabels=false}{hyperref}
\usepackage{hyperref}
\hypersetup{%
    colorlinks=true,%
    urlcolor=ETHg,%
    linkcolor=ETHc,%
    citecolor=ETHb,%
    breaklinks=true,%
    pageanchor=true,%
}

\usepackage[activate={true,nocompatibility},final,tracking=true,%
            kerning=true,spacing=true,factor=1100,stretch=10,shrink=10]%
            {microtype}
\microtypecontext{spacing=nonfrench}

\SetTracking{encoding={*}, shape=sc}{0}

\useshorthands*{"}
\defineshorthand{"-}{\babelhyphen{hard}}
\defineshorthand{"=}{\allowbreak}

\makeatletter
\DeclareRobustCommand{\varname}[1]{\begingroup\newmcodes@\mathit{#1}\endgroup}
\makeatother

\newcommand\algo[1]{\textsc{#1}}
\newcommand\algobf[1]{\textsc{\bfseries #1}}
\newcommand\repro[2]{\texttt{repro\allowbreak <#1,\allowbreak #2>}}
\newcommand\pair[2]{$\langle\varname{#1},\allowbreak\varname{#2}\rangle$}
\newcommand\ulp{\mathrm{ulp}}
\newcommand\ufp{\mathrm{ufp}}
\newcommand\rd{\mathrm{rd}}

\pgfplotscreateplotcyclelist{bar plot colors}{%
    {ETHe,fill=ETHe!60!white,pattern color=ETHe},
    {ETHa,fill=ETHa!60!white,pattern color=ETHa},
    {ETHc,fill=ETHc!60!white,pattern color=ETHc},
    {ETHb,fill=ETHb!60!white,pattern color=ETHb},
    {ETHd,fill=ETHd!60!white,pattern color=ETHd}
}
\pgfplotscreateplotcyclelist{bar plot patterns}{%
    {},
    {postaction={pattern=north east lines}}
}

\title{\mbox{\hspace{-.3em}Reproducible Floating-Point Aggregation in RDBMSs}\\%
    {\Large [Extended Version]}}

\author{
    Ingo Müller$^{\text{1}*}\protect\footnote{hello world}$ \hspace{1.5cm}
    Andrea Arteaga$^\text{2}$ \hspace{1.5cm}
    Torsten Hoefler$^\text{1}$ \hspace{1.5cm}
    Gustavo Alonso$^\text{1}$
    \vspace{1.6mm}\\
    \fontsize{10}{10}\selectfont\rmfamily\itshape
        $^\text{1}$Systems Group, Dept. of Computer Science, ETH Zurich \\
    \fontsize{9}{9}\selectfont\ttfamily\upshape
       \{ingo.mueller, torsten.hoefler, alonso\}@inf.ethz.ch
    \vspace{1.2mm}\\
    \fontsize{10}{10}\selectfont\rmfamily\itshape
       $^\text{2}$Federal Institute of Meteorology and Climatology MeteoSwiss \\
    \fontsize{9}{9}\selectfont\ttfamily\upshape
       andrea.arteaga@meteoswiss.ch
}

\begin{document} 

\maketitle

{\let\thefootnote\relax\footnotetext{$^*$%
                 Part of this work was carried out
                 while this author was working part-time
                 at Oracle Labs, Zurich.}}

\begin{abstract}
Industry-grade database systems are expected
to produce the same result
if the same query is repeatedly run on the same input.
However, the numerous sources of non-determinism in modern systems
make reproducible results difficult to achieve.
This is particularly true if floating"-point numbers are involved,
where the order of the operations affects the final result.

As part of a larger effort to extend database engines
with data representations more suitable
for machine learning and scientific applications,
in this paper we explore the problem of making relational \algo{GroupBy}
over floating"-point formats \emph{bit"-reproducible},
i.e., ensuring any execution of the operator produces
the same result up to every single bit.
To that aim, we first propose a numeric data type
that can be used as drop-in replacement for other number formats
and is---unlike standard floating"-point formats---associative.
We use this data type
to make state-of-the-art \algo{GroupBy} operators reproducible,
but this approach incurs a slowdown between \factorrange{4}{12}
compared to the same operator using conventional database number formats.
We thus explore how to modify existing \algo{GroupBy} algorithms
to make them bit"-reproducible \emph{and} efficient.
By using vectorized summation on batches and
carefully balancing batch size, cache footprint, and preprocessing costs,
we are able to reduce the slowdown due to reproducibility
to a factor between \factorrange{1.9}{2.4} of aggregation in isolation
and to a mere \SI{2.7}{\percent} of end-to-end query performance
even on aggregation-intensive queries in MonetDB.
We thereby provide a solid basis for supporting
more reproducible operations directly in relational engines.

This document is an extended version
of an article currently in print for the proceedings of ICDE'18
with the same title and by the same authors.
The main additions are more implementation details and experiments.
\end{abstract}

\section{Introduction} 

The continued progress of all areas of computer science
has led to digitization and automation of many everyday processes.
In particular, algorithms are responsible for making decisions
in an increasing number of commonplace situations%
~\cite{FederalTradeCommission2016}.
Consequently---and understandably---,
society has started to demand accountability
from the algorithms by which it is affected.
For example, the General Data Protection Regulation
of the European Union~\cite{EuropeanParliament2016}
has recently given the ``right to explanation''
to individuals affected by automated decision-making.
Similarly, the ACM published
a \emph{Statement on Algorithmic Transparency and Accountability}%
~\cite{ACMUSPublicPolicyCouncil2017}
including explainability as a principle
that algorithmic decision-making should follow,
a problem studied for example in the context of computational journalism%
~\cite{Diakopoulos2016,Diakopoulos2016a}.

In this paper we take steps towards improving
explainability of today's data processing systems,
namely the reproducibility of algorithms based on floating-point arithmetic.
This problem was brought to our attention by several of our industry partners.
In addition to more classical use-cases like
debugging, testing, certification, and redundant computations,
where reproducibility can be helpful or necessary~\cite{Chiang2013,Blelloch2012},
they observe that many users, in particular non-experts,
are confused by non-reproducible
or, in general, non-predictable behavior.

Today's data management systems often become non-reproducible
if floating-point arithmetic is used.
The problem with floating-point arithmetic is that,
unlike arithmetic on real numbers, it is not associative,
so the order in which operations are carried out
may change their outcome~\cite{Goldberg1991}.
The order of operations, in turn, may be affected by a large list of mechanisms:
For example, concurrent execution of multiple threads
may be non-deterministic,
the number of processing elements
may influence how the work is split into sub-tasks,
and the data storage layer may physically reorder data for a number of reasons.
As soon as floating-point numbers are involved,
most of today's systems do not follow
the principle of \emph{data independence},
which demands that changes on the physical level
shall not have an impact on the result of queries.

\begin{algorithm}
  \begin{center}
    \vspace{-.1cm}
    \begin{minted}[mathescape,
                   gobble=2,
                   bgcolor=white,
                   framesep=2mm,
                   fontsize=\small,
          ]{sql}
  CREATE TABLE R        (i int, f float);
  INSERT INTO  R VALUES (1,     2.5e-16);
  INSERT INTO  R VALUES (2,     0.999999999999999);
  INSERT INTO  R VALUES (3,     2.5e-16);
  SELECT SUM(f) FROM R; -- Returns 0.999999999999999
  UPDATE R SET i = i + 1 WHERE i = 2;
  -- 'f' is unchanged, but rows are physically reordered
  SELECT SUM(f) FROM R; -- Returns 1.0!
    \end{minted}
    \vspace{-.8cm}
  \end{center}
  \caption{Example of non-reproducible SQL query.}
  \label{alg:mvcc-affects-sql-query}
  \vspace{-.2cm}
\end{algorithm}

Algorithm~\ref{alg:mvcc-affects-sql-query} illustrates
how the order of records in the storage layer
may affect query results in a subtle and potentially surprising way.
The situation shown was produced on a fresh installation of PostgreSQL 9.5.1.
The same query summing up three floating-point numbers
returns two different results before and after
the update of an unrelated attribute.
Since, internally, the update is implemented as the creation of a new record
and the masking of the old one,
the physical order is different in the two queries,
which, consequently, yields two different results
(differing on all digits of the decimal representation).

One may be tempted to brush away the problem of non"-reproducibility
with the argument that the underlying rounding errors are rather small
and can, hence, be ignored.
However, these small errors may still lead
to very different outcomes for individual records,
which are hard to explain to the affected individual.
For instance, the \algo{GroupBy} query
of Algorithm~\ref{alg:mvcc-affects-sql-query}
extended with a \mintinline{sql}{HAVING SUM(f) >= 1} clause
could end up returning specific records in some runs but not in others.

Such misclassifications can affect applications in obvious ways:
We ran PageRank on different permutations
of a small web graph with \SI{900}{\kilo\nothing} pages.%
\footnote{\url{https://snap.stanford.edu/data/web-Google.html}}
We observed that, from one run to the next,
the ranks of about 10-20 pages
would be \emph{different enough to swap ranks with another page}.

In this paper we show
how to make \algo{GroupBy} aggregation
using \algo{Sum} reproducible.
This essentially solves the problem for SQL:
With a reproducible aggregate function for floating-point \algo{Sum},
\emph{all} aggregate functions in SQL can be made reproducible as well,
including non-standard ones such as \algo{Variance},
\algo{StdDev}, and some statistical functions,
all of which can be computed using \algo{Sum}.%
\footnote{Aggregate functions such as \algo{Min}, \algo{Median},
          \algo{Count}, \algo{Rank}, \algo{Percentile}, and similar,
          but also functions such as \algo{ListAgg} and \algo{Collect},
          do not need floating-point arithmetic.
          The remaining functions offered by the Oracle database
          (see \url{https://docs.oracle.com/database/121/SQLRF/functions003.htm})
          can be computed with \algo{Sum}.}
Furthermore, many projections are intrinsically reproducible%
\footnote{If arithmetic expressions are computed in their entirety
          only once all operands are available,
          then executing them always in the same order is trivial.
          However, if expressions are broken up
          and partially pushed through joins,
          then their execution order may depend on the join order,
          which may change even if the logical input has not changed.
          However, we believe that this should be solved
          on the level of the optimizer,
          which is out of the scope of this paper.}
and window functions can either be solved with our approach,
define an execution order,
or they are not reproducible even with integers.%
\footnote{Window clauses without sliding frame
          can be executed as aggregations with \emph{GroupBy}.
          Window clauses with \algo{OrderBy} clause have a definite order
          and are therefore intrinsically reproducible
          and without \algo{OrderBy} clause,
          the result may change even with integers.}
Finally, \algo{GroupBy} aggregation is not only used
in relational database systems,
but in virtually every data processing system
(possibly under a different name
including \algo{Reduce} or \algo{ReduceByKey}),
to which our results apply as well.

We start with proposing a format for floating-point numbers
that is---unlike formats typically supported by hardware---associative.
The format can be implemented in software
and builds on techniques from high-performance computing (HPC)%
~\cite{Arteaga2014,Demmel2013,Demmel2015}.
The key to this approach is to anticipate rounding errors
by subtracting lower-order terms from each value
before it is added to the aggregate of its group.

While this makes it possible to make any algorithm on floating-point numbers
bit-reproducible with little to no modification,
it comes at a high price:
We show that state-of-the-art \algo{GroupBy} operators
become about \factorrange{4}{12} slower using this approach,
depending on the desired precision.
The challenge is, hence,
to keep the overhead of the additional calculations at an acceptable level.
However, the tuning techniques used in HPC
work for the sum of a single vector,
while in a SQL \algo{GroupBy},
there is potentially a very large number of sums involved.
This is not compatible with known data processing techniques,
which usually aggregate input tuples as early as possible
instead of physically ``grouping'' them.
State-of-the-art aggregation algorithms used with our data type,
hence, switch between the summation of different groups
\emph{for every input tuple},
which explains the high overhead.

To remedy this problem, we design a novel \algo{GroupBy} algorithm
based on a concept we call \emph{summation buffer}.
The main idea is to buffer input values for each group
and delay their aggregation
until it can be done efficiently for the whole buffer.
As we need a summation buffer for every group,
the number of groups that we can process efficiently at the same time
is limited by how many buffers we can keep in cache.
We thus tune the summation routine to small buffer sizes
and use highly-tuned partitioning routines as preprocessing.
This reduces the slowdown of aggregations due to reproducibility
to a factor between \factorrange{1.9}{2.4}
over non-reproducible aggregation on built-in floating"-point numbers,
depending on the number of groups in the input.
Integration into a real system, MonetDB~\cite{Boncz2008},
shows that we can bring the overhead of end-to-end query performance
down to as little as \SI{2.7}{\percent}.
Since our implementation
hides almost all computations behind memory accesses,
we can even increase accuracy with minimal additional cost
and, hence, as a side effect provide higher accuracy than IEEE numbers
at essentially the same price,
which is crucial in many scientific applications.

To summarize, the paper makes the following contributions:
\begin{itemize}
  \item We propose a highly tuned algorithm
    for reproducible summation of floating-point numbers
    using SIMD instructions
    (Section~\ref{sec:reproducible-summation}).
  \item We show how state-of-the-art algorithms
    for aggregation with \algo{GroupBy}
    can be made bit"-reproducible and more accurate
    with relatively little effort
    if compromises in performance can be made
    (Section~\ref{sec:naive-integration}).
  \item We design a novel grouping algorithm that improves upon this approach,
    which reduces the slowdown of reproducibility
    to a \factorrange{1.9}{2.4}
    (Section~\ref{sec:buffered-aggregation}).
  \item We show the trade-offs offered by the different algorithms
    in extensive experiments
    and quantify their impact on end-to-end query performance in a real system
    (Section~\ref{sec:evaluation}).
\end{itemize}

\section{Problem Definition} 
\label{sec:basics}

We start by illustrating
the cause of non-reproducibility of floating-point summation
and by discussing potential solutions for bit"-reproducibility,
which, unfortunately, either do not actually solve the problem
or have prohibitive costs.

\subsection{Reproducible Floating-Point
            Aggregation with \algobf{GroupBy}}

For the purpose of this paper,
we define aggregation with \algo{GroupBy} as the operation
that turns a sequence of \pair{key}{value} pairs
into the \pair{key}{aggregate} pairs
where each key of the input occurs exactly once in the output
and the aggregate stored with a key
is equal to the sum of all values with that key in the input.
We say that it runs on floating-point numbers
if the $\varname{value}$ fields of the input pairs are floating-point numbers.
An aggregation algorithm is bit"-reproducible, or reproducible for short,
if the $\varname{aggregate}$ of each group
has exactly the same bit pattern for any execution.

\subsection{Floating-Point Numbers and Associativity}
\label{sec:floats}

Floating-point values are numbers of the form $x = M \cdot 2^E$,
where $M \in \left[ 1, 2 \right)$ is called \emph{mantissa}
and $E \in \left\{ E_{\varname{min}}, E_{\varname{max}} \right\}$
is the \emph{exponent}.
As the number of relevant bits $m$ in the mantissa
as well as the exponent are finite,
only a finite subset of real numbers can be represented exactly.
Hence, a rounding function $\rd$ is required
in order to map real numbers to representable floating"-point values.
This includes the results of arithmetic expressions,
which may not be representable even if the operands are.
For example, the floating-point sum of two floating-point numbers $a$ and $b$
is defined as $a \oplus b = \rd(a+b)$.

To understand why this can be a problem,
consider the numbers $a = b = 1.01_2 \cdot 2^0$ and $c = 1.11_2 \cdot 2^1$
in a toy format for floating-point numbers
with a mantissa of $m=2$ (given as binary number)
and truncation for $\rd$.
To compute the sum of the three numbers,
we can compute $(a \oplus b) \oplus c = \rd(\rd(a+b) + c)$.
Since $\rd(a+b) = \rd(1.01\bm{0}_2 \cdot 2^1) = 1.01_2 \cdot 2^1$
and $\rd(1.01_2 \cdot 2^1 + c) = \rd(1.10\bm{0}_2 \cdot 2^2) = 1.10_2 \cdot 2^2$,
no rounding errors occur and the sum is accurate.
However, we can compute the sum as well as
$a \oplus (b \oplus c) = \rd(a + \rd(b+c))$.
In this case $\rd(b+c) = \rd(1.00\bm{11}_2 \cdot 2^2) = 1.00_2 \cdot 2^2$
and $\rd(a + 1.00_2 \cdot 2^2) = \rd(1.01\bm{01}_2 \cdot 2^2) = 1.01_2 \cdot 2^2$,
so rounding errors occur in both operations (typeset in bold).
Note that the \emph{sum} of the rounding errors is $1.00_2 \cdot 2^0$,
which could be added to $a \oplus (b \oplus c)$ without rounding error.

Rounding errors are larger
if the exponents of the two summands are different.
Therefore, if we compute the sum of many numbers,
the rounding error incurred during the addition of a particular input value
depends on the current value of the accumulator,
which depends on the order of execution.
Furthermore, even though each error is small,
their sum may be big enough to change the final result.

The problem also occurs in the most common floating-point formats,
which are the ones defined by the IEEE-754 standard~\cite{Zuras2008IEEE}
(even if the absolute error is obviously smaller
due to the higher precision than our toy format)
and which we use throughout this paper.

\subsection{Non-Solutions for Reproducibility}
\label{sec:non-solutions}

We now discuss a number of naive approaches
for making aggregation with \algo{GroupBy} reproducible,
but which do not provide satisfactory solutions to the problem.

\begin{description}[nosep,labelindent=\parindent,leftmargin=0pt]
  \item[Higher precision.]
    Using a truncated or rounded result
    produced by operations with a higher floating"-point precision
    (i.e., using doubles instead of floats)
    is not sufficient,
    as it does not make it more reproducible:
    Even tiny rounding errors can make significant bits flip
    (such as from $0.999999\ldots$ to $1$).

\itodo{Ce: there is also cuDNN, which does the same thing}
  \item[Deterministic order of operations.]
    It is possible to make the order of the operations deterministic.
    For linear algebra,
    the cuBLAS library~\cite{NVIDIACorporation2016}
    and the Intel Math Kernel Library~\cite{Rosenquist2012}
    follow this approach.
    However, this does not solve the entire problem for database systems,
    which aim for \emph{data independence} as discussed above.
    In addition to the example given in the introduction,
    the physical order of the input may also change
    due to compression, data placement on distributed machines,
    backup and restore operations, and other mechanisms,
    which in turn changes the order of operations.
    The only way to make the order of the operations deterministic
    is thus to use a static and deterministic schedule
    \emph{and} to sort the input,%
    \footnote{Provably so: Visiting a set of elements in a fixed order
              solves the \algo{ProximateNeighbor} problem,
              which is as hard as \algo{Sorting}~\cite{Jacob2014}.}
    which may be
    more than an order of magnitude slower~\cite[Figure 5.8]{Balkesen2014}
    than state-of-the-art \algo{Aggregation} algorithms based on hashing.

  \item[Fixed-point arithmetic.]
    In traditional workloads,
    it is often possible to use fixed-point arithmetic for fractional numbers
    (also called \emph{binary scaling}
    or \emph{decimal"-scaled binaries} if the scale has base ten).
    This is the case if all input numbers
    are integer multiple of some common denominator
    and come from similar orders of magnitude,
    such as all values in a \emph{salary} field are multiples of $1\cent$
    and range between some thousand and some million dollars.
    Operations can then be executed using integer operations internally,
    which are cheap to execute and reproducible most of the time.%
    \footnote{Summing integers may experience overflows,
              which can lead to non"-reproducible results
              if the values have mixed signs and
              depending on how they are handled.
              If overflows are to be prevented in software,
              this may incur a slowdown
              as high as \factor{3} as well~\cite{Dietz2012}.}
    However, in many modern data processing applications,
    these assumptions do not apply:
    values from some domains, such as measurements or scientific data,
    cannot be expressed as multiple of some smallest unit
    and the values of different orders of magnitude
    such as those handled in machine learning and other scientific applications
    require a floating"-point representation.

\itodo{ingo: I am not sure about the following, in particular Oracle Numbers \\
       ingo: I am sure enough for the first submission.
       I will check again for the final one. \\
       Jana: are high-precision numbers/Oracle numbers actually reproducible?}

  \item[Arbitrary-precision operations.]
    It is possible to push the limits of fixed-point arithmetic by using
    high-precision or even arbitrary-precision operations in software.
    Examples implementing this approach include
    the GNU MPFR Library~\cite{MPFR},
    the BigDecimal class of Java~\cite{JavaBigDecimal},
    and numeric data types offered by some database systems (such as PostgreSQL).
    However, this not only requires many hardware instructions
    for each arithmetic operation,
    but also variable-width storage,
    which is much more difficult to handle than the fixed-width built-in types.
    Similar arguments apply to \emph{unums}~\cite{JohnGustafson2015}.

%
%

\end{description}

\section{Reproducible Accurate Sum} 
\label{sec:reproducible-summation}

In this section, we explain an algorithm
that solves the problem of reproducible floating-point summation
where all inputs are summed up to produce a \emph{single} number
(i.e., aggregation \emph{without} grouping).
For the sake of exposition, we develop the algorithm
in three iterations of increasing completeness.
In a fourth iteration, we show how to speed up this algorithm
using vector instructions.
Table~\ref{tbl:symbols} summarizes the variables and function names we use.

\begin{table}
    \centering
    \begin{tabular}{l@{\hspace{2em}}l}
        \toprule
        \textbf{Name} & \textbf{Meaning} \\
        \midrule
        $n$ & Number of input values  \\
        $L$ & Number of levels  \\
        $V$ & Size of vector register \\
        $W$ & Bit width of error-free transformation \\
        $\varname{NB}$ & Block size between carry-bit operations \\
        $m$ & Size of the mantissa \\
        $f$ & Exponent of the first transformation level \\
        \midrule
        $b_i$ & Input value  \\
        $q^{(l)}_i$ & Contribution of $b_i$ at level $l$ \\
        $r^{(l)}_i$ & Remainder of $b_i$ at level $l$ \\
        $S^{(l)}$ & Running sum at level $l$ \\
        $C^{(l)}$ & Carry-bit count at level $l$ \\
        $Q^{(l)}$ & Sum of the contributions at level $l$ \\
        \midrule
        $\ufp(x)$ & Unit in the first place \\
        $\ulp(x)$ & Unit in the last place \\
        $\rd(x)$ & Rounding function \\
        $x \oplus y  $ & $\equiv \rd(x+y)$, floating"-point sum of $x$ and $y$  \\
        $x \ominus y $ & $\equiv \rd(x-y)$, floating"-point subtraction of $x$ and $y$  \\
        \bottomrule
    \end{tabular}
    \caption{\label{tbl:symbols}
           Summary of the parameters, variables, and functions used
           in Section~\ref{sec:reproducible-summation}.}
\end{table}

\subsection{Definitions}

Two numbers related to any floating"-point number $x$
are of importance in this section: $\ufp(x)$ and $\ulp(x)$.
They were first defined by \textcite{Goldberg1991}.
$\ufp(x)$ designates the \emph{unit in the first place},
i.e., the numeric value of the first bit in the mantissa.
If $x = M \cdot 2^E$ as above, then $\ufp(x) = 2^{E}$.
All floating"-point numbers
with the same exponent as $x$
have an absolute value in $\left[\ufp(x),2\cdot\ufp(x)\right)$.
Similarly, $\ulp(x)$ designates the \emph{unit in the last place},
i.e., the value of the last bit in the mantissa.
For $x = M \cdot 2^E$, $\ulp(x) = 2^{E-m}$.
This value represents the difference between $x$ and its
closest representable values.

\subsection{Error-Free Transformation}

Our reproducible summation algorithm is
based on the observation that the floating"-point sum of two numbers $a$ and $b$
can be performed \emph{exactly} if the one with the smaller absolute value has
sufficiently many zeroes at the end of its mantissa.
To understand when this is the case, let us define $a$ and $b$ as
integer multiples of the same power of two: $a := \alpha_a \cdot 2^p, b :=
\alpha_b \cdot 2^p$, where $\alpha_a, \alpha_b$, and $p$ are integer numbers, and
$|a| > |b|$, WLOG\@.
If
the values $|\alpha_a|, |\alpha_b|$, and $|\alpha_a + \alpha_b|$ do not exceed
$2^m$, then $a$, $b$, and $a+b$
can be represented in the floating"-point format and, consequently,
the floating"-point sum is exact: $a \oplus b = a + b$.
In other words, the sum is exact if these three numbers can have a (possibly
denormalized) floating"-point representation with the same exponent and without
losing information.

As an example, the values $a:=26.046875$ and $b:=2.8125$ can be represented
exactly with an 11-bit mantissa (which corresponds to IEEE half"-precision, where $m = 10$). Also, they
can be represented as $\alpha_a \cdot 2^p$ and $\alpha_b \cdot 2^p$, with
$\alpha_a = 1667$, $\alpha_b = 180$, and $p = -6$ and their sum is
$a \oplus b = a + b = (\alpha_a + \alpha_b) \cdot 2^p = 28.859375$. It can be computed
exactly with this format because said conditions are met.

Let us now consider any representable $a$ and $b$ with $|a| > |b|$. One can
split $b$ as
$b = q + r$ so that $q$ is an integer multiple of $\ulp(a)$,
namely $q := (a \oplus b) \ominus a$ and $r := b \ominus q$.
The sum $a \oplus q$ can be computed exactly because the three aforesaid conditions are
met between $a$ and $q$, with $2^p = \ulp(a)$. Also, $b$ can be recovered
exactly through $q+r = q \oplus b$. This procedure, named \emph{error-free
transformation}, was defined by Ogita et al.~\cite{Ogita2005}.

For instance, $a = 1.010_2 \cdot 2^0, b = 1.101_2 \cdot 2^{-2}$. $q$ is computed
as $(a \oplus b) \ominus b = 1.101_2 \cdot 2^0$, while $r = b \ominus q = 1_2
\cdot 2^{-5}$.

\begin{figure}
    \centering
    \includegraphics[width=\columnwidth]{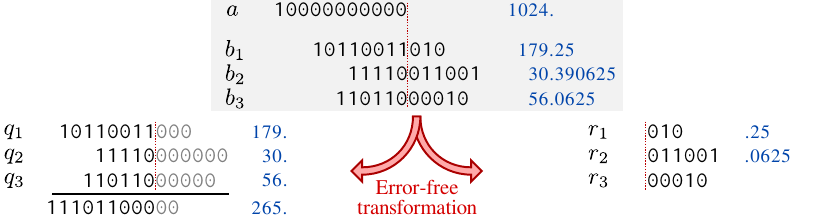}
    \caption{\label{fig:error-free-transformation}Error"-free transformation.}
\end{figure}

We can now perform an order-independent summation of a sequence of values
$b_1, b_2, b_3$ by
applying the same error-free transformation to all values $b_i$.
Figure~\ref{fig:error-free-transformation} illustrates the idea.
The transformation produces $q_i := (a \oplus b_i) \ominus a$,
$r_i := b_i \ominus q_i$.
We call the values $q_i$ and $r_i$
\emph{contributions} and \emph{remainders}, respectively, of the input values $b_i$
and the value $a$ the \emph{extractor} of the error-free transformations.
Since all contributions $q_i$ are integer multiples
of the same power of two, and that their sum is representable with the same
format,
their floating"-point summation is free of rounding error
and
thus order-independent: $q_1 \oplus q_2 \oplus q_3 =
q_1 + q_2 + q_3 = 265$.

While this procedure solves the problem of order-independent
bit-reproducibility, it has two problems:
First, it only works under the assumption that the
absolute value of every intermediate result of this summation, including the
final result, is strictly bounded by $2 \cdot \mathrm{ufp}(a)$.
An obvious, yet suboptimal solution would be to do two passes over the data, the first one
computing the maximum absolute value for which all assumptions are
fulfilled, the second one for the actual summation.
Second, the sum is inaccurate because relevant parts of the input
values, namely the remainders, are discarded.
We present an algorithm that solves both problems.

\subsection{Accurate Reproducible Scalar Summation}

In HPC, the problem of summing up long vectors of numbers
has been studied in detail.
In this section, we explain \algo{RSum}~\cite{Demmel2015}.
In the subsequent section,
we discuss why it does not work well with SQL \algo{GroupBy}.

In order to address the problem of accuracy, we recall that the error-free
transformation produces two outputs: the \emph{contribution} $q$ and the
\emph{remainder} $r$. In the previous section, we made use of the contributions
to obtain an order-independent summation. Now we make use of the remainders
to improve the accuracy of this summation: we perform an error-free
transformation on the remainder, this time using the smaller extractor $a^{(2)} := 2^{-W}
\cdot a$ (with $W \in \mathbb{N}\setminus\{0\})$.
We thus obtain the second-level contribution
and remainder of each input value: $q_i^{(2)} := (a^{(2)} \oplus r_i)
\ominus a^{(2)}, r_i^{(2)} := r_i \ominus q_i^{(2)}$. These second-level
contributions can be summed up to obtain a second-level result: $Q^{(1)} :=
\sum_{i=1}^{n} q_i, Q^{(2)} := \sum_{i=1}^{n} q_i^{(2)}$.
As we show in Section~\ref{sec:tuning-summation},
the final result $Q :=
Q^{(1)} \oplus Q^{(2)}$ is of comparable accuracy as a
standard, non-reproducible floating"-point summation. If higher accuracy is
needed, an arbitrary number of levels $L$ can be used. The value $W$ expressing the
logarithm of the ratio of two consecutive extractors $a^{(l)},
a^{(l+1)}$ is bounded by $m-2$ and it affects the result (the
higher, the more accurate) and the cost (the higher, the slower) of the
algorithm. Good choices are 18 and 40 for single and double precision
respectively and we use these values in this work.

So far, the extractors $a$ are never assumed to be powers of two. The example in
Figure~\ref{fig:error-free-transformation} shows a power of two as the
extractor, but this does not necessarily have to be the case. The only important
factor is that the exponent of the extractor never changes, nor do the
intermediate results of the error-free transformation. For this reason,
the role of the error-free extractor is taken by the running sums $S^{(l)}$ in the
algorithm. The running sum will never change its exponent, as is explained later
in this section.

We start with the values $S^{(l)} = 1.5 \cdot
2^{f-(l-1) \cdot W}$.
The value $f$ can be chosen arbitrarily, as long as the resulting extractor is
large enough for the transformation of the first value $b_1$, i.e.,
$f > \log_2 \left| b_1 \right| + m - W + 1$.
Each
input value $b_i$ is transformed using $S^{(1)},"=S^{(2)}, \dots,"=S^{(L)}$ as
extractors. The resulting contributions $q_i^{(1)},"=q_i^{(2)},"=
\dots, q_i^{(L)}$ are added to $S^{(1)},"=S^{(2)}, \dots,"=S^{(L)}$ respectively.
For $|b_i| \ge 2^{W-1} \cdot \ulp\big(S^{(1)}\big)$, the first level is not large
enough to contain its contribution. In this case, the last level $S^{(L)}$ is
discarded, all other levels are demoted (e.g., the first-level sum becomes
the second-level sum, $S^{(1)} := S^{(2)}$, and the new first-level sum is initialized
to $S^{(1)} := 1.5 \cdot 2^W \cdot \ufp\big(S^{(2)}\big)$. This ensures that all
input values can be included in the summation without breaking the assumptions
for reproducible results, also avoiding the need for a first pass to
find the maximum absolute value in the input.

In order to avoid that the running sums $S^{(l)}$ change their exponent (which would
affect the error-free transformation), a check is performed before its usage.
$S^{(l)}$ is \emph{usable}, if it lies in the range $\left[ 1.5 \cdot
\ufp\left( S^{(l)} \right), 1.75 \cdot \ufp\left(S^{(l)}\right) \right)$. If it does
not, a multiple of $0.25 \cdot \ufp(S^{(l)})$ is added to or removed from
it, and the corresponding value is subtracted from or added to the carry-bit
counter $C^{(l)}$, which is initialized to 0. For instance, if $S^{(l)} = 1.84
\cdot \ufp(S^{(l)})$, then $1 \cdot \left( 0.25 \cdot
\ufp(S^{(l)}) \right)$ is subtracted from it, so that the running sum after
this operation is $S^{(l)} = 1.59 \cdot \ufp(S^{(l)})$; the value 1 is
added to $C^{(l)}$. The complete state of the summation is given by the
running sums $S^{(l)}$ and the corresponding carry-bit counts $C^{(l)}$.

\begin{algorithm}
    \begin{algorithmic}[1]
        \STATE Load state $S^{(l)}$, $C^{(l)} \;\; \forall 1 \le l \le L$
        \FOR{$i=1$ \textbf{to} $n$}
            \STATE \label{loc:begin-checklevel} $\rhd$ Check extractor
            validity, update levels if needed: necessary
            \WHILE{$|b_i| \ge 2^{W-1} \cdot \ulp \left( S^{(1)} \right)$}
                \FOR{$l=L$ \textbf{to} $2$}
                    \STATE $S^{(l)} \leftarrow S^{(l-1)}; \quad C^{(l)} \leftarrow C^{(l-1)}$
                \ENDFOR
                \STATE $S^{(1)} \leftarrow 1.5 \cdot 2^W \cdot \ufp(S^{(2)}); \quad C^{(1)} \leftarrow 0$
            \ENDWHILE \label{loc:end-checklevel}
            \STATE \label{loc:begin-transform} $\rhd$ Load and transform value $b_i$, update $S^{(l)}$:
            \STATE $r^{(0)}_i \leftarrow b_i$
            \FOR{$l=1$ \textbf{to} $L$}
                \STATE $q^{(l)}_i \leftarrow \left( r^{(l-1)}_i \oplus S^{(l)} \right) \ominus S^{(l)}$
                \STATE $S^{(l)} \leftarrow S^{(l)} \oplus q^{(l)}_i$
                \STATE $r^{(l)}_i \leftarrow r^{(l-1)}_i \ominus q^{(l)}_i$
            \ENDFOR \label{loc:end-transform}
            \STATE \label{loc:begin-carry} $\rhd$ Carry-bit propagation:
            \FOR{$l=1$ \textbf{to} $L$}
                \STATE Find $d \in \mathbb{Z}$ s.t. $S^{(l)} \ominus d \cdot 0.25 \cdot \ufp\left( S^{(l)} \right)
                       \in \left[ 1.5 \cdot \ufp \left( S^{(l)} \right), 1.75 \cdot \ufp \left( S^{(l)} \right) \right)$
                \STATE $S^{(l)} \leftarrow S^{(l)} \ominus d \cdot 0.25 \cdot \ufp\left( S^{(l)} \right)$
                \STATE $C^{(l)} \leftarrow C^{(l)} \oplus d$
            \ENDFOR \label{loc:end-carry}
        \ENDFOR
        \STATE Store state $S^{(l)}$, $C^{(l)} \;\; \forall 1 \le l \le L$
    \end{algorithmic}

    \caption{\label{alg:rsum-scalar}
             \algo{RSum Scalar}.}

\end{algorithm}

Algorithm~\ref{alg:rsum-scalar} summarizes this procedure.
In this version of the algorithm we assume that the summation state has been
initialized. A number of input values are added to this summation state, and
its state is stored to main memory again, so that the summation can
be resumed later. In order to finalize the summation, the following sum has to
be performed:
\begin{equation}
    Q := \sum_{l=1}^L \left( \left( S^{(l)} \ominus 1.5 \cdot \ufp(S^{(l)}) \right)
    \oplus 0.25 \cdot \ufp\left( S^{(l)}\right) \cdot C^{(l)} \right)
\end{equation}

This sum is not order-independent, thus a predefined order has to be
imposed. In order to avoid cancellation, we perform it in reverse order, i.e., we
start from the last level.

\begin{figure}
    \centering
    \includegraphics[width=\columnwidth]{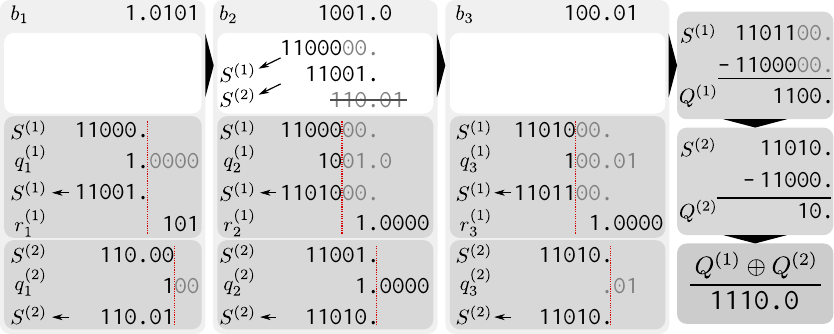}
    \caption{\label{fig:rsum-example}Application of the \algo{Rsum} algorithm on
     three values.}
\end{figure}

\begin{figure}
    \centering
    \includegraphics[width=\columnwidth]{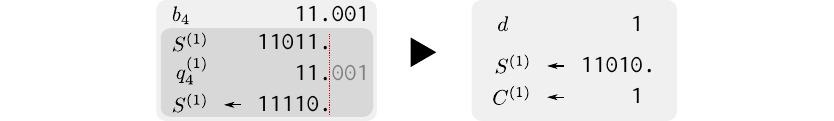}
    \caption{\label{fig:carry-example}Example of a carry-bit propagation.}
\end{figure}

Figure~\ref{fig:rsum-example} shows the application of
Algorithm~\ref{alg:rsum-scalar} on the values $b_1 = 1.3125$, $b_2 = 9$,
and $b_3 = 4.25$, with the format defined by $m = 4$,
the parameter $W = 2$, the first extractor chosen with $f = 4$, and two
extraction levels. The figure uses only binary digits.
In the first iteration,
the value $b_1$ is added to the first-level running sum $S^{(1)}$
(incrementing it by the contribution $q_1^{(1)}$).
The remainder $r_1^{(1)}$ is added to the second-level running sum $S^{(2)}$.
In the second iteration,
$|b_2| \ge 2^{W-1} \cdot \ulp \left( S^{(1)} \right)$, thus triggering an
adjustment of the levels, shown in the white box: The second-level sum is
discarded, the first-level sum is moved to the second level, and a new extractor
is set as first level. Then, the extraction is performed normally. The third
value does not trigger the level adjustment. Finally,
the sum for each level is computed
($Q^{(1)}$ and $Q^{(2)}$) and these values are
summed to give the final result $1110_2 = 14$. $C^{(l)}$ variables are never shown
in this example because their value is always zero, as
$S^{(l)} \in \left[ 1.5 \cdot \ufp \left( S^{(l)} \right),
1.75 \cdot \left( S^{(l)} \right) \right)$ at all times.
Figure~\ref{fig:carry-example} illustrates a carry-bit operation on one level
when the running sum $S^{(1)}$ has the value $11011_2$, the carry-bit counter
$C^{(1)} = 0$, and the value $b_4 = 3.125$ is processed:
after the summation the running sum exceeds
$1.75 \cdot \ufp \left( S^{(l)} \right)$. The adjusting number is found to be $d
= 1$, and the sum and the carry-bit counter are modified accordingly.

\subsection{Vectorization of the Summation Algorithm}

\itodo{Ingo: I think we should give a better motivation why we do this.
       This could be a forward reference to the evaluation
       that gives the peak speed-up we achieve by vectorization.
       However, I am not sure what the speed-up really is:
       it looks like in some cases,
       vectorization makes the algorithm 8x faster.
       Was the original summation routine really that bad?}

\algo{RSum}~\cite{Demmel2015} was originally introduced in a MIMD
context, where each process performs the full summation of the local data and
the results are finally summed up globally using \texttt{MPI\_Reduce}.
As a first step to make it suitable to \algo{GroupBy},
we propose a SIMD variant of \algo{RSum}.
In this variant, the running sum of each level is represented in the
registers as a tuple of values $\left\langle S^{(l)}_1, \dots, S^{(l)}_V
\right\rangle$, where
$V$ is the width of the register (e.g., for double-precision values on AVX
architectures, $V = 4$). Similarly, the carry-bit counts are represented by a
tuple of $V$ elements and $V$ input values are transformed and added to the running
sums concurrently.
Moreover, a tiling optimization is performed: the extractor validity and the
carry"-bit propagation are performed just once every $\varname{NB}$ iterations.
This is bounded by $\varname{NB} \le 2^{-m-W-1}$\cite{Demmel2015}.

The summation state does not change format: one running sum and one carry-bit
count per level are stored in main memory. When loading a summation state
from memory into the registers, the first element of the registers is set to
the value read from memory (thus, e.g., $S^{(1)}_1 = S^{(1)}$), while the other
elements of each register are initialized to $1.5 \cdot \ufp(S^{(l)})$
for running sums and to 0 for carry-bit counts. A horizontal summation has to
be performed at the end of the algorithm when storing the state to the main
memory: The resulting running sum and carry-bit count of each level are:
\begin{align}
    S^{(l)} &:= 1.5 \cdot \ufp(S^{(l)}_1) \oplus \sum_{v=1}^V
        \left( S^{(l)}_v \ominus 1.5 \cdot \ufp(S^{(l)}_v) \right),\\[-.8em]
    C^{(l)} &:= \sum_{v=1}^V C^{(l)}_v
\end{align}
with order-independent sums.
Algorithm~\ref{alg:rsum-simd} summarizes this procedure. For brevity and clarity,
parts of the algorithm constitute references
to the equivalent lines of Algorithm~\ref{alg:rsum-scalar}.

\begin{algorithm}
    \begin{algorithmic}[1]
        \STATE Load state $S^{(l)}_1 \leftarrow S^{(l)}; \;\;
                           C^{(l)}_1 \leftarrow C^{(l)}
                           \;\; \forall 1 \le l \le L$
        \STATE $S^{(l)}_v \leftarrow 1.5 \cdot \ufp\big( S^{(l)} \big); \;
                C^{(l)}_v \leftarrow 0 \; \forall 2 \le v \le V, \, 1 \le l \le L$
        \FOR{$i=1$ \textbf{t}o $n$, \textbf{increment by} $V \cdot \varname{NB}$}
            \STATE $\rhd$ Check $\max_{i \le j < i+V \cdot \varname{NB}} |b_j|$,
                   update levels if necessary. See Algorithm~\ref{alg:rsum-scalar},
                   lines~\ref{loc:begin-checklevel}-\ref{loc:end-checklevel}
            \FOR{$j=i$ \textbf{to} $i+V \cdot \varname{NB}$, \textbf{increment by} $V$}
                \STATE $\rhd$ Load and transform values
                       $\left\langle b_j, \ldots, b_{j+V-1} \right \rangle$,
                       update $\left\langle
                       S^{(l)}_1,\ldots,S^{(l)}_V \right\rangle$.\\
                       See Algorithm~\ref{alg:rsum-scalar},
                       lines~\ref{loc:begin-transform}-\ref{loc:end-transform}
            \ENDFOR
            \STATE $\rhd$ Carry-bit propagation.
            See Algorithm~\ref{alg:rsum-scalar},
            lines~\ref{loc:begin-carry}-\ref{loc:end-carry}
        \ENDFOR
        \STATE $\rhd$ Horizontal summation
        \FOR{$l=1$ \textbf{to} $L$}
            \STATE $S^{(l)} \leftarrow 1.5 \cdot \ufp(S^{(l)}_1) \oplus \sum_{v=1}^V
                   \left( S^{(l)}_v \ominus 1.5 \cdot \ufp(S^{(l)}_v) \right)$
            \STATE $C^{(l)} \leftarrow \sum_{v=1}^V \left\langle
            C^{(l)}_1,\ldots,C^{(l)}_V \right\rangle$
        \ENDFOR
        \STATE Store state $S^{(l)}$, $C^{(l)} \;\; \forall 1 \le l \le L$
    \end{algorithmic}

    \caption{\algo{RSum Simd}.}
    \label{alg:rsum-simd}
\end{algorithm}

\section{A Reproducible Floating-Point Type}
\label{sec:naive-integration}

As a first solution
for reproducible floating-point aggregation with \algo{GroupBy},
we propose a data type
that can be used as drop-in replacement
for intermediate aggregates of floating-point numbers
in any state-of-the-art aggregation algorithm
with little to no modification.%
\footnote{The only arithmetic operation that this type supports is addition,
          so in a real system it would most likely be
          an internal type of the execution layer
          not exposed to the user.}
We base this type on the reproducible summation algorithm
from the previous section:
It simply consists of an $\bigl\langle\vec{S},\vec{C}\bigr\rangle$ pair,
where the symbols
$\vec{S} = \bigl\langle S^{(1)},\allowbreak\ldots,\allowbreak S^{(L)}\bigr\rangle$ and
$\vec{C} = \bigl\langle C^{(1)},\allowbreak\ldots,\allowbreak C^{(L)}\bigr\rangle$
are, respectively, the $L$ levels of the running sum and carry-bit counter
as introduced in Section~\ref{sec:reproducible-summation}.
In languages such as C++, we can implement this data type
as a class with member variables \texttt{S[L]} and \texttt{C[L]}
and overload its \texttt{operator+=}
for summation with scalars and instances of that type.
We refer to this data type as \repro{ScalarT}{L},
where \texttt{ScalarT} is either float or double.

\itodo{ingo: we have to say how to implement these functions!
       maybe we can present the math of Section~\ref{sec:reproducible-summation}
       as algorithms and refer to those? \\
       Gustavo: yes!}

\begin{figure}
   \includegraphics[width=\columnwidth]{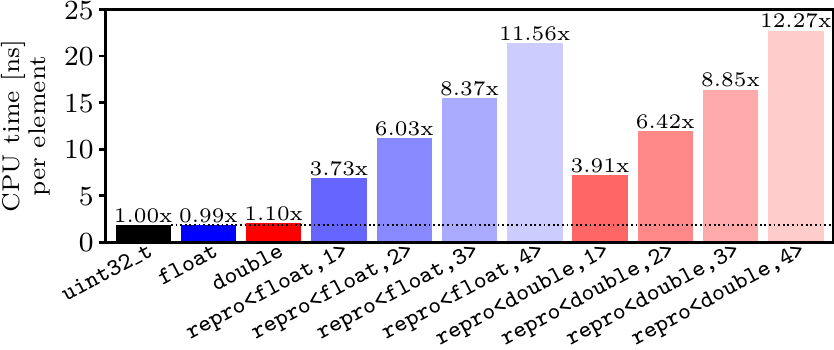}
   \caption{\algo{HashAggregation}
   with different reproducible data types and 16 groups.}
   \label{fig:naive-slowdown}
\end{figure}

Figure~\ref{fig:naive-slowdown} shows the performance
of a start-of-the-art \algo{HashAggregation} algorithm%
\footnote{The experimental setup as described in Section~\ref{sec:eval-setup}.}
instantiated with different variants of the \texttt{repro} data type.
This algorithms looks up the aggregate of the corresponding group
in a hash table using the $\varname{key}$ field of the input pair
and adds the $\varname{value}$ field to that aggregate.
We choose the small number of 16 groups to eliminate
effects not related to the data types themselves
(such as cache effects or pre-processing costs).
As the plot shows, the algorithm is between \factorrange{4}{12} slower
with the reproducible data types than with integers or IEEE floats
and the more so the more levels of summation we use
(i.e., the higher the precision).
This is not surprising considering the computational overhead:
in the \texttt{operator+=} of reproducible types,
each level of summation requires about 12 floating"-point operations
and 4 load and store instructions,
while the \texttt{operator+=} of standard data types
only requires a single one.
Finally, there is virtually no difference between single and double precision.
This is due to the fact that the algorithm is heavily compute bound
and the latency of most instructions does not depend on the operand width.

We have learned two things from this section.
First, it requires relatively little development effort
to make a large class of algorithms for aggregation with \algo{GroupBy}
bit"-reproducible on floating"-point numbers.
Second, this approach comes at a high price:
If we want to match the precision of IEEE floats,
we need the types with $\text{\texttt{L}} = 2$,
which, in the situation shown above,
makes the algorithm more than 6 times slower.
In the next section, we show a more involved algorithm
that improves the slowdown to between \factorrange{1.9}{2.4}
for any \repro{ScalarT}{L}.

\vspace{-1mm}
{\sloppy
\section{Aggregation with Summation Buffers} 
\label{sec:buffered-aggregation}}

In this section, we present a novel algorithm
for aggregation with \algo{GroupBy}
that achieves bit"-reproducibility on floating"-point numbers efficiently.
We first describe the main idea, summation buffers,
which allows to use our efficient, vectorized summation routine,
and then incorporate that
into a state-of-the-art \algo{Aggregation} algorithm.

\subsection{Summation Buffers}
\label{sec:accumulation-buffer-overview}

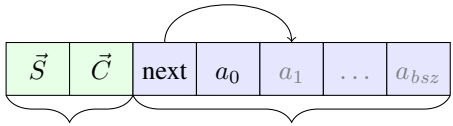
\begin{figure}
  \begin{center}
    \newlength\mctextheight
\setlength\mctextheight{\heightof{$\vec{C}\vec{S}$}}
\begin{tikzpicture}[
      every node/.style={draw,minimum width=2.4em,minimum height=2em,
            text height=\mctextheight,text depth=depth("y$a_0$")},
      key/.style={fill=red!10!white},
      val/.style={fill=green!10!white},
      buf/.style={fill=blue!10!white},
      empty/.style={text=gray}
    ]
  \draw
    (0,0)       node [val] (state-begin) {$\vec{S}$}
    ++(2.4em,0) node [val] (state-end) {$\vec{C}$}
    ++(2.4em,0) node [buf] (c) {next}
    ++(2.4em,0) node [buf] (buffer-begin) {$a_0$}
    ++(2.4em,0) node [buf,empty] (a) {$a_1$}
    ++(2.4em,0) node [buf,empty] {$\ldots$}
    ++(2.4em,0) node [buf,empty] (buffer-end) {$a_{\varname{bsz}}$}
    ;
  \node [fit=(state-begin)(state-end), inner sep=-.5pt, draw=none] (state) {};
  \node [fit=(c)(buffer-begin)(buffer-end), inner sep=-.5pt, draw=none] (buffer) {};
  
  \draw [->] (c) to [in=90,out=90] (a);
  \draw [decorate,decoration={brace,amplitude=10pt}]
    (state.south east) -- (state.south west)
    node [draw=none,midway,yshift=-0.85em,xshift=-1em,anchor=north] {%
\begin{BVerbatim}
repro<ScalarT,L>
\end{BVerbatim}
};
  \draw [decorate,decoration={brace,amplitude=10pt}]
    (buffer.south east) -- (buffer.south west)
    node [draw=none,midway,yshift=-0.85em,anchor=north] {accumulation buffer};
\end{tikzpicture}
  \end{center}
  \vspace{-5mm}
  \caption{Memory layout of an intermediate aggregate.}
  \label{fig:accumulation-buffers}
\end{figure}

The main idea for batching the aggregation of input values
can be used in any \algo{Aggregation} algorithm:
we store a reproducible float along with a buffer of input values
as intermediate aggregates, which we call \emph{summation buffer}.
A summation buffer consists of an array of input values
and the offset of the next free slot in the array (``next'').
The layout of intermediate aggregates is thus
as shown in Figure~\ref{fig:accumulation-buffers}.

For example, the textbook \algo{HashAggregation}~\cite{Muller2016}
with summation buffers works as follows:
Whenever we process a \pair{key}{value} pair,
we first use the key to lookup the entry of the group in the hash table
and then use the offset to append the value to the buffer of the group
(incrementing the offset accordingly).
Only when a buffer is full, we aggregate the buffered values
and reset the offset to 0, i.e., to the beginning of the buffer.
This allows us to use our vectorized summation algorithm
\algo{RSum Simd} (Algorithm~\ref{alg:rsum-simd}),
which is much more efficient
than the per-element summation from the previous section.
In languages such as C++, we can implement this as new data type again,
where the summation operators contain the logic just described,
and use this new data type
in any existing \algo{Aggregation} algorithm transparently.

One important tuning parameter is now the buffer size ($\varname{bsz}$).
On the one hand, the larger the buffer is,
the better the constant costs
associated with a call to the summation algorithm can be amortized.
On the other hand, the larger each buffer is,
the larger is also the cache footprint of the algorithm,
which may decrease performance significantly.
The rest of this section shows how to design an algorithm
that makes the best trade-off in all situations.

\subsection{High-Level Algorithm Structure}
\label{sec:algorithm-structure}

The overall structure of our \algo{Aggregation} algorithm
is illustrated in Algorithm~\ref{alg:partition-and-aggregate}:
We partition the input on the hash value of the keys
(Line~\ref{loc:paa-partition}),
which can be done very efficiently on modern hardware%
~\cite{Boncz2008,Schuhknecht2015,Muller2015}.
Since all input records for a particular group
are copied into the same partition,
each partition can be processed independently,
which we do using \algo{HashAggregation}
(Lines~\ref{loc:paa-hashagg-start} to \ref{loc:paa-hashagg-end}).
Finally, we combine the intermediate results
into a single hash table shared among all threads
(Lines~\ref{loc:paa-merge-start} to \ref{loc:paa-merge-end}).
We call this algorithm \algo{PartitionAndAggregate}.

\begin{algorithm}
  \begin{algorithmic}[1]
    \STATE $\text{partitions} \leftarrow \text{\textsc{ParallelPartition}(input, $\varname{key}$, $F=f^d$)}$
        \label{loc:paa-partition}
    \FOR{\textbf{each} p \textbf{in} partitions \textbf{with index} i \textbf{parallel}}
      \label{loc:paa-hashagg-start}
      \STATE $\text{privateTables[i]} \leftarrow \text{\textsc{HashAggregation}(p)}$
      \label{loc:paa-hashagg-body}
    \ENDFOR \label{loc:paa-hashagg-end}
    \FOR{\textbf{each} t \textbf{in} privateTables \textbf{parallel}}
      \label{loc:paa-merge-start}
      \FOR{\textbf{each} \pair{key}{value} \textbf{in} t}
        \STATE $\text{sharedTable}[\varname{key}] \mathrel{+}= \varname{value}$
        \label{loc:paa-merge-body}
      \ENDFOR
    \ENDFOR
    \label{loc:paa-merge-end}
  \end{algorithmic}
  \caption{\algo{PartitionAndAggregate}.}
  \label{alg:partition-and-aggregate}
\end{algorithm}

The partitioning effectively divides the number of groups per partition
and, hence, the cache footprint of \algo{HashAggregation}
by the partitioning fan"-out $F$,
which may out-weigh the additional costs for partitioning.
Depending on the number of groups in the input,
a larger or smaller fan"-out is needed
in order to fit the working set of \algo{HashAggregation} into the cache.
For a small number of groups, no partitioning may be required.
In this case, i.e., if $F=1$,
\algo{ParallelPartition} is a no-op that forwards its input.
Since modern hardware can run \algo{Partitioning} efficiently
only up to a certain fan"-out%
~\cite{Boncz2008,Schuhknecht2015,Muller2015},
we implementing it recursively using zero or more levels of partitioning
i.e., we partition with $F=f^d$ for $f=256$ and $d=0,1,\ldots$.

All phases of the algorithm can be fully parallelized:
The partitioning routine called in Line~\ref{loc:paa-partition}
can be parallelized by splitting the input in an arbitrary way
(e.g., into equally-sized chunks, using a work queue, or using work stealing)
and logically concatenating the corresponding output partitions
produced by different threads.
After the partitioning, each thread gets a subset of the partitions,
which it aggregates into a private hash table
independently of the other threads
(Lines~\ref{loc:paa-hashagg-start} to \ref{loc:paa-hashagg-end}).
With some care, the subsequent transfer to the shared hash table
(Lines~\ref{loc:paa-merge-start} to \ref{loc:paa-merge-end})
can be implemented without synchronization
since the threads work on non-overlapping subsets of its content.
If no partitioning needs to be done, i.e., if $F=1$,
the shared hash table needs some form of synchronization such as locks.
However, since in this case there are only few groups
and each of them only appears once in the hash table of each thread,
this last phase takes a negligible amount of time,
so the overhead of locking is acceptable.

In order to make this algorithm reproducible,
we use summation buffers as the data type
for the intermediate aggregates produced by \algo{HashAggregation}
in Line~\ref{loc:paa-hashagg-body}.
In the process of aggregating its share of the input or its partitions,
each thread calls \texttt{operator+=(ScalarT)}
on the appropriate intermediate aggregates,
which makes it effectively alternate between probing the hash table
in order to append input values at the end of their corresponding buffers
and summing up the content of buffers as they become full.
The shared hash table used in Line~\ref{loc:paa-merge-body}
has aggregates of type \repro{ScalarT}{L},
i.e., it does not use summation buffers
and \texttt{operator+=(\allowbreak repro\allowbreak <ScalarT,\allowbreak L>)} is used
for merging the intermediate aggregates of the different threads.

The careful reader may wonder
why the data of a particular partition
is first aggregated into a private hash table
and then transferred into a part of the shared hash table
that the thread has exclusive access to.
It seems possible to save the transfer
by aggregating into summation buffers in the shared hash table directly.
However, this would have several disadvantages:
\begin{enumerate*}[label=(\arabic*)]
  \item for all but almost distinct inputs,
    the transfer is negligible as argued earlier,
    so there is no need to speed up this phase,
  \item the result would consist of summation buffers,
    which take up more space than needed, and
  \item to finalize the computation,
    we need to iterate over the results anyway
    in order to flush the buffers.
\end{enumerate*}

\subsection{Tuning Buffer Size and Partitioning Depth}
\label{sec:parameter-tuning}

\algo{PartitionAndAggregate} with summation buffers has thus two parameters
that influence its cache footprint:
the size of the summation buffers $\varname{bsz}$
and the partitioning depth $d$.
We now show how to choose these two parameters.

We first determine the size of the summation buffers
given a fixed partitioning depth.
Since the access to the summation buffers follows the random pattern
given by the hash values of the keys of the input records,
the cache footprint of the algorithm consists of the size of the hash table,
which we can quantify as
$n_{\varname{groups}} \cdot \allowbreak
    \text{\texttt{sizeof}}\allowbreak(\texttt{ScalarT}) \allowbreak
    \cdot \varname{bsz}$,
where $n_{\varname{groups}}$ is the number of groups in the input.
The buffers should be as large as possible
in order to amortize constant costs of calling the summation routine.
We thus set the buffer size such that they use the entire cache,
which is given by the following equation:
\begin{gather}
  \varname{bsz} = \min
    \begin{cases}
      \lceil |\varname{cache}| /
        \left( n_{\varname{groups}} / F\cdot
               \text{\texttt{sizeof}}(\texttt{ScalarT}) \right) \rceil \\
        \varname{bsz}_{\varname{max}},
    \end{cases}
    \raisetag{.46cm}
  \label{eqn:2pass-bsz}
\end{gather}
where $|\varname{cache}|$ is the size of the last-level cache
corresponding to one thread, $\varname{bsz}_{\varname{max}}$
the largest buffer size available in the system,
and $F$ the partitioning fan"-out.

The optimal number of levels of partitioning $d$
depends on the number of groups:
It must be large enough to reduce the number of groups per partition to a point
where the subsequent, final level of aggregation can be done in cache.
It should not be larger, otherwise, the partitioning has no benefit
and its execution only constitutes overhead.
In earlier work~\cite{Muller2015}, we
propose to select this depth adaptively:
start with a private hash table of fixed size;
while the number of groups is lower than the threshold,
process all input this way;
if and when the threshold is crossed,
add a level of partitioning and recurse.
This has virtually no overhead,
so the resulting runtime essentially corresponds
to the optimal partitioning depth
for any given input.
Since the adaptation mechanism is orthogonal to the topic of this paper
and incorporation into our algorithm is only a matter of implementation time,
we simply determine the optimal number of levels offline
and use that in the remainder of this paper.

%

\subsection{System Integration}

We envision the integration of our algorithm into real systems
either as a ``fix'' for \algo{Sum} on floating-point numbers
or as an alternate aggregate function
\algo{RSum}($\langle\varname{expression}\rangle$, $\varname{L}$),
which would give the user control on the desired precision.

\section{Experimental Evaluation} 
\label{sec:evaluation}


In this section, we show micro-benchmarks
that justify design decisions taken in the previous sections,
evaluate the performance of our algorithms experimentally,
and quantify their impact on end-to-end query performance.

\subsection{Experimental Setup}
\label{sec:eval-setup}

We run the experiments on a system with \SI{256}{GiB} RAM
and two Intel Xeon E5-2630 v3 CPUs,
which belong to the Haswell-EP product line.
The CPUs have 8 physical cores each clocked at \SI{2.4}{\giga\hertz},
each with private first and second-level data caches
of size \SI{32}{\kibi\byte} and \SI{256}{\kibi\byte}, respectively,
as well as a \SI{20}{\mebi\byte} last-level cache shared among all cores.
The system runs Debian 8 (jessie) with a Linux kernel v3.18.14.
HyperThreading and frequency scaling are switched off.

Unless otherwise mentioned, we use $n = 2^{30}$
\pair{key}{value} pairs as input,
where the key is of type \texttt{uint32\_t}
and the type of the value is as follows:
It is of type \texttt{ScalarT}
if we say that an algorithm runs on \repro{ScalarT}{L}
(i.e., it is of type \texttt{float} or \texttt{double})
and of type \texttt{DECIMAL(p)}
if we say that the algorithms runs on one of these types.
We implement the \texttt{DECIMAL} types as built-in integers
of size 32, 64, and 128 bit%
\footnote{For 128-bit integers, we use the \texttt{\_\_int128} type
          available in recent versions of GCC on our CPU.}
for $\text{\texttt{p}}=9,19,38$, respectively,
which is a typical way to implement them.
The keys are drawn uniformly at random
from the range $[0,n_{\varname{groups}})$.
Due to the nature of random distributions,
this means that there are actually
less than $n_{\varname{groups}}$ groups in the input
if $n_{\varname{groups}} \approx n$.
We omit experiments on other data distributions
as known techniques to handle data skew~\cite{Cieslewicz2007,Muller2015}
are orthogonal to the topic of this paper
and can be included into our algorithms.
All presented numbers are averages of ten identical runs,
among which we observed low variance.%
\footnote{The relative standard deviation
          is mostly below \SI{1}{\percent}
          and never above \SI{5}{\percent}.}
We express the runtime as
$\text{``CPU time per element''} = T \cdot P / n$,
where $T$ is the total running time,
$P=8$ the number of processing elements,
and $n$ the number of input elements,
which simplifies comparison across different machines.

We implemented our algorithms in C++,
which we compile with GCC version 4.9.2.
Sporadic tests with newer compilers did not bring significant improvements,
probably because we manually force optimizations
such as function inlining, loop unrolling, and vectorization
wherever beneficial.
In order to reduce the impact of the operating system,
we pin each thread to a different core of a single socket
and allocate and initialize all memory on that socket before the experiment.
We refrain from using several sockets in order to avoid NUMA effects,
again referring to the fact
that NUMA-optimizations are orthogonal to reproducibility.

In experiments not shown in this paper,
we compared our baseline implementation to that of \textcite{Cieslewicz2007}.
Our implementation is somewhat faster,
mainly because we use \algo{IdentityHashing} instead of multiplicative hashing.
This is not unrealistic in column stores,
where dense ranges are common due to domain encoding.
Using a real hash function would make
all our algorithms slower by the same constant
and thus result in even lower relative overhead of reproducibility.
Our implementation of \algo{PartitionAndAggregate}
is up to 4 times faster than that of \textcite{Cieslewicz2007}
because we use the highly-tuned partitioning routine used in other work%
~\cite{Boncz2008,Polychroniou2014,Schuhknecht2015}.
Back-of-the-envelope calculations suggest that we achieve the same performance
as the implementations used in \cite{Muller2015}, as well,
thereby ensuring our baseline for \algo{GroupBy} matches the state of the art.

We also ran experiments with an implementation of \algo{SortAggregation}
based on the highly-tuned sorting routines of \textcite{Balkesen2014},
which, as discussed above, can be used to make aggregation reproducible
by bringing the input in a deterministic order.
Even on integers or built-in floats,
this algorithm needs a CPU time per element of over \SI{60}{\nano\second},
which is \SI{20}{\x} more than our algorithm in the best case
and at least \SI{3}{\x} more in any case
where $n/n_{\varname{groups}} < 2^6$,
so we did not pursue this approach further.

\subsection{Vectorized Summation Algorithm}
\label{sec:tuning-summation}

We first evaluate accuracy and performance of our summation routine
presented in Section~\ref{sec:reproducible-summation}.

\subsubsection{Accuracy}

The accuracy of our summation routine \algo{RSum Simd}
(Algorithm~\ref{alg:rsum-simd})
depends on the number of levels of running sums and carry-bit counters, $L$.
To quantify the accuracy of our routine, we compare its absolute error
with that of the non-reproducible summation of conventional floats.
According to \textcite{Demmel2013}, the latter error can be bounded by:
\vspace{-4mm}
\begin{equation}
    e_{conv} := (n-1) \cdot \varepsilon \cdot \sum_{i=1}^n |b_i|,
\end{equation}
where $\varepsilon$ represents a machine constant~\cite{Goldberg1991}
and $b_i$, $i=1..N$, represent the summands.
The error of our routine can be bounded the following expression,
which is due to \textcite{Demmel2015} and the same for their and our algorithm:
\begin{equation}
    e_{\algo{Rsum Simd}} := n \cdot 2^{(1-L) \cdot W - 1} \cdot \max_{1 \le i \le n} |b_i|,
\end{equation}
where $f$ and $W$ are the exponent of the first extractor used
and the ratio of two consecutive extractors, respectively.

\begin{table}
    \centering
    \begin{tabular}{@{}l<{\hspace{.7em}}>{\hspace{-.7em}}ll@{}>{\hspace{.6em}~}l>{\hspace{-.7em}}ll@{}}
        \toprule
                            & \multicolumn{2}{c}{$n=10^3$}                   && \multicolumn{2}{c}{$n=10^6$}               \\
                            \cmidrule{2-3} \cmidrule{5-6}
                            & $\mathrm{U}[1, 2)$     & $\mathrm{Exp}(1)$     && $\mathrm{U}[1, 2)$   & $\mathrm{Exp}(1)$   \\
        \midrule
        Conventional        & $1.7 \cdot 10^{-10}$   & $1.1 \cdot 10^{-10}$  && $1.7 \cdot 10^{-4}$  & $1.1 \cdot 10^{-4}$ \\
        \algo{Rsum} ($L=1$) & $1.0 \cdot 10^{3}$     & $1.1 \cdot 10^{4}$    && $1.0 \cdot 10^{6}$   & $1.1 \cdot 10^{7}$  \\
        \algo{Rsum} ($L=2$) & $9.1 \cdot 10^{-10}$   & $1.0 \cdot 10^{-8}$   && $9.1 \cdot 10^{-7}$  & $1.0 \cdot 10^{-5}$ \\
        \algo{Rsum} ($L=3$) & $8.3 \cdot 10^{-22}$   & $9.1 \cdot 10^{-21}$  && $8.3 \cdot 10^{-19}$ & $9.1 \cdot 10^{-18}$\\
        \bottomrule
    \end{tabular}
    \vspace{2mm}
    \caption{\label{tbl:sum-errors}
           Maximum absolute error of conventional and reproducible summation
           algorithms in double precision.}
    \vspace{-6mm}
\end{table}

To quantify these expressions in a simple way,
we consider the summation of random arrays,
with uniformly-distributed values in the range $[1, 2)$ and
exponentially-distributed values with $\lambda = 1$, with varying number of
values $n$. With the latter distribution, the probability of an input set with
$10^6$ values to contain a value larger than 22 is lower than $0.03\%$, thus we
choose 22 as the maximum expected input value and we use it to give a reasonable
error bound.
Table~\ref{tbl:sum-errors} shows the expected values of the
error bounds for the different algorithms and different distributions in double precision.
The error bound for the single-level reproducible summation can be surprisingly
large. The reason for such large bounds is the low control on the magnitude of
the levels the algorithm has. The
largest extractor used for the summation could have a much larger magnitude than
the result (up to $2^{W-1}$ times larger). In this unfortunate case, only one
significant bit of the result is kept by the first summation level. If this is
the only level used, the uncertainty on the result is as large as the result
itself. In many cases, the one-level summation can deliver reasonably accurate
results and, for large input sizes, its accuracy can be comparable to the
conventional summation. Nevertheless, it gives no guarantee of accurate results
beyond the error bounds listed in the table. All error bounds for the
reproducible algorithm are up to $2^{W-1}$ times more pessimistic than the actual
error of the summations.

\begin{figure}
    \centering
    \includegraphics{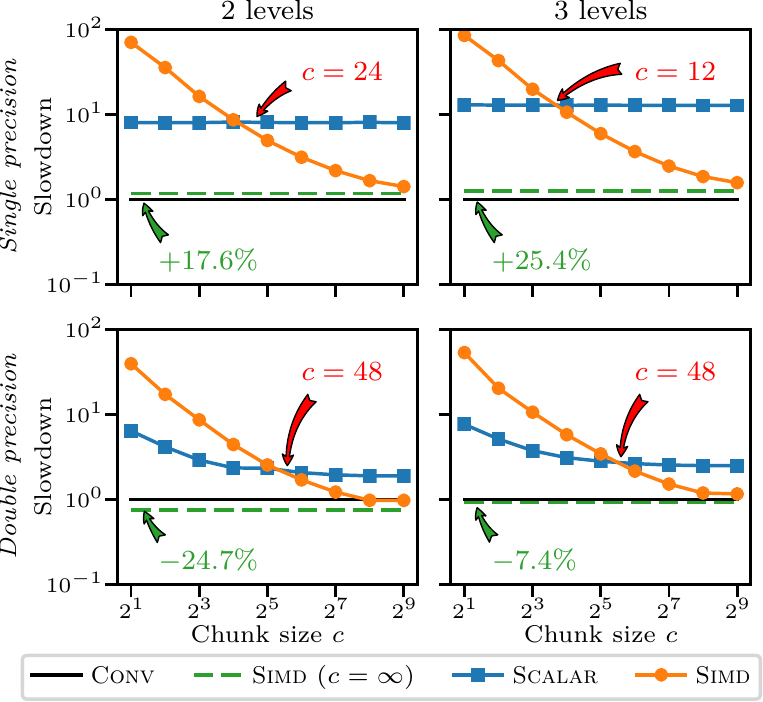}
    \caption{\label{fig:chunks}
        Relative performance of \algo{RSum} algorithms
        compared to a conventional sum using \texttt{std::accumulate} (\algo{Conv}).
    }
    \vspace{-1mm}
\end{figure}

\textbf{Conclusion:}
Our summation routine \algo{Rsum Simd} with $L=2$ has comparable accuracy
as conventional floating-point summation
and achieves much higher accuracy with $L>2$.

\medskip

\subsubsection{Performance}

We also measure the performance
of several variants of the reproducible summation algorithm (\algo{RSum})
for summing up a large array of random numbers.
Figure~\ref{fig:chunks} shows the result.
For \algo{RSum Scalar} and \algo{RSum Simd}
(Algorithms~\ref{alg:rsum-scalar} and~\ref{alg:rsum-simd}, respectively),
we sum up the input in chunks of $c$ values for various values of $c$,
i.e., we call the respective algorithm for each chunk of $c$ values.
This mimics the pattern
of how the summation algorithms are used
in the aggregation algorithms of Section~\ref{sec:buffered-aggregation},
where they switch between the summation of inputs of different groups.
For brevity, the algorithms are simply called
\algo{Scalar} and \algo{Simd} in the figure.
We plot the performance in terms of slowdown
compared to a conventional summation algorithm on the same input,
which is called \algo{Conv} in the figure
and implemented as a single call to \texttt{std::accumulate}.
Finally, as ``lower bound'',
we also plot the performance of a single call to \algo{RSum Simd},
which we call \algo{Simd} ($c=\infty$).

As the figure shows,
\algo{RSum Simd} is slower than \algo{RSum Scalar} for small chunk sizes,
but faster for large chunk sizes.
For small chunk sizes, it suffers from a higher start-up overheads
because the state it loads and stores from memory into registers and back
is a factor $V$ times larger than that of the scalar version.
For large chunk sizes, vectorization pays off.
The cross-over point (annotated in red)
is somewhere between $c=12$ and $c=48$
depending on the number of levels $L$ and the precision.
As the chunk size reaches $c=512$,
the start-up overhead of the multiple calls to \algo{RSum Simd} is amortized
and the performance reaches that of \algo{Simd} ($c=\infty$),
which consists of a single call.
At this point, \algo{RSum Simd}
is at most \SI{25}{\percent} slower than \algo{Conv}
and even somewhat faster in case of double precision.
We attribute this to the fact
that the compiler is not able to fully vectorize \texttt{std::accumulate},
whereas \algo{RSum Simd} ($c=\infty$) is optimized
to the point that it is memory-bound.

\textbf{Conclusion:}
Our summation routine is faster, the larger the buffer size $\varname{bsz}$,
as constant start-up costs can be amortized better.
For $\varname{bsz} \geq 2^6$, \algo{Simd} is always better than \algo{Scalar}
and for $\varname{bsz} \geq 2^9$ or earlier,
the difference to the maximum throughput is negligible,
where the routine is memory-bound.

\subsection{Aggregation without Summation Buffers}
\label{sec:eval-naive-integration}

We first present experiments of
unmodified state-of-the-art aggregation algorithms,
i.e., we do not use aggregation buffers.
We compare two categories of data types:
\begin{enumerate*}[label=(\arabic*)]
  \item The reproducible \repro{ScalarT}{L} types
    presented in Section~\ref{sec:naive-integration},
    which corresponds to making \algo{Aggregation} reproducible
    with minimal development effort, and
  \item \texttt{DECIMAL(p)} with various precisions \texttt{p}
    (where \texttt{p} is the number of decimal digits),
    which may be a good enough alternative for some applications.
\end{enumerate*}
We emphasize that, as discussed in Section~\ref{sec:non-solutions},
the \texttt{DECIMAL} types are not flexible enough
for many modern applications,
which cannot determine the scale of the involved numbers statically
and thus need a \emph{floating}-point representation.
We still include them as a reference point.
We also ran the algorithms with built-in floating"-point types,
but observed exactly the same performance
as \texttt{DECIMAL(9)} and \texttt{DECIMAL(18)}
for \texttt{float} and \texttt{double}, respectively,
so we omit them in the plots shown below.

In order to determine how much partitioning is needed as pre-processing,
we use the same procedure as described below
in Section~\ref{sec:eval-buffered-hashagg}
for each data type separately:
while for built-in types, using one and two levels pays off
starting from $2^{16}$ and $2^{25}$ groups, respectively,
the thresholds vary somewhat for the reproducible data types.
They range from $2^{14}$ to $2^{16}$ and $2^{20}$ to $2^{24}$, respectively.
This is due to varying amounts of overhead of \texttt{operator+=},
which exceeds the costs of \algo{Partitioning}
sometimes sooner, sometimes later,
but can be easily accommodated for in practice.

\begin{figure}
    \includegraphics[width=\columnwidth]{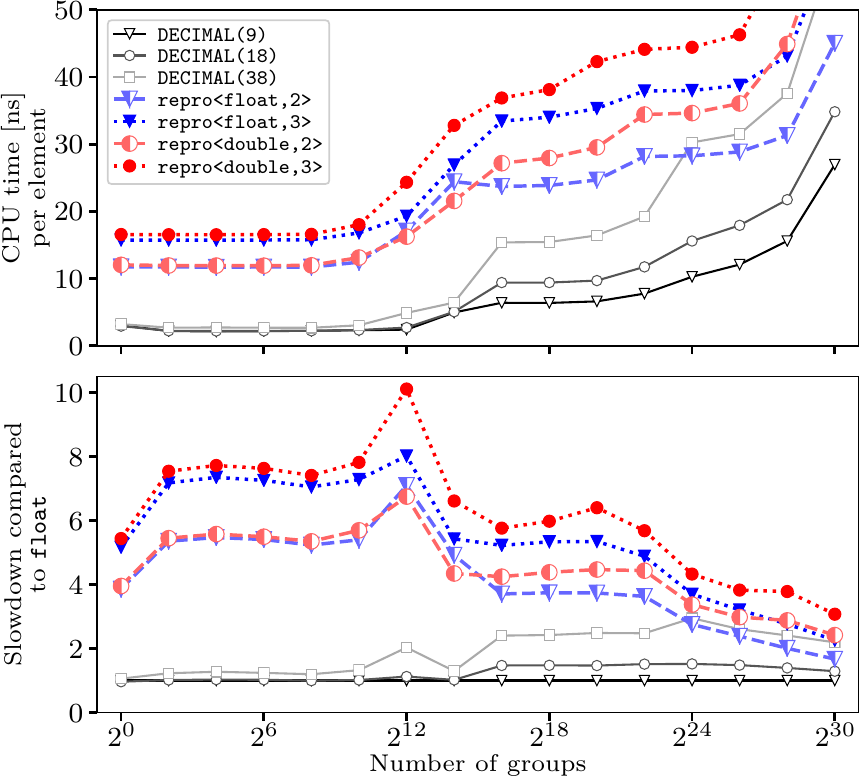}
    \caption{\algo{PartitionAndAggregate} on various \repro{ScalarT}{L}
             \emph{without} summation buffers
             compared to the same algorithm on \texttt{float}/\texttt{DECIMAL(8)}.}
    \label{fig:unbuffered-comparison}
    \vspace{-1mm}
\end{figure}

Figure~\ref{fig:unbuffered-comparison} summarizes our findings.
For better readability, we only show the results
for $\text{\texttt{L}}=2$ and $\text{\texttt{L}}=3$,
which are the most interesting configurations in practice.
As in the experiment shown in Figure~\ref{fig:naive-slowdown},
the different levels are about equidistant from each other,
which gives an idea of the omitted data points.

\begin{figure*}
    \centering
    \includegraphics[width=\textwidth]{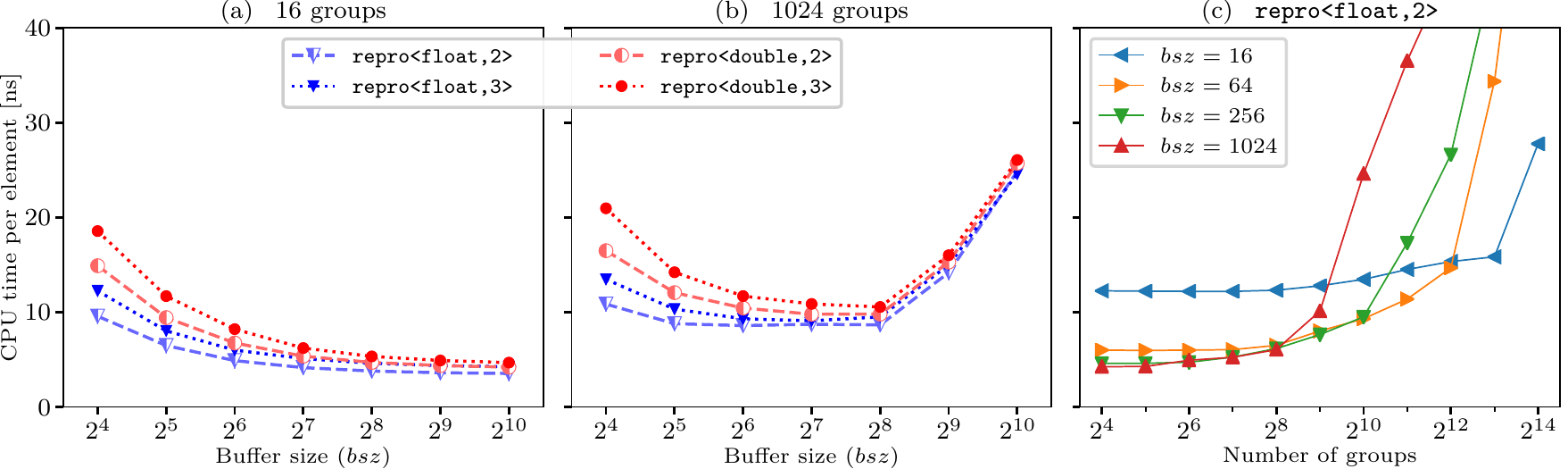}
    \caption{Impact of the buffer size on \algo{PartitionAndAggregate} with $d=0$.}
    \label{fig:buffersize-pass1}
    \vspace{-1mm}
\end{figure*}

The upper diagram shows that the runtime
for all data types follows the expected pattern:
fast, in-cache processing for small numbers of groups
and more and more overhead for the partitioning
as the number of groups increases,
each ``step'' corresponding to an additional level of partitioning.
For \texttt{DECIMAL(p)}, the steps are higher the larger \texttt{p},
which is due to the higher memory traffic for wider data types in that phase.
As the keys in the input become more and more distinct
(for most data types at around $2^{27}$ groups in the $2^{30}$ input records),
the costs of evicting the final result back to RAM becomes noticeable.
Furthermore, the larger \texttt{L}, i.e., the higher the accuracy,
the longer the runtime of \repro{ScalarT}{L} by quite a large difference,
slightly more so for doubles than for floats.
Finally, for \texttt{DECIMAL(p)}, the steps due to partitioning
are higher the larger \texttt{p},
which is due to the higher memory traffic for wider data types in that phase.

The lower diagram in Figure~\ref{fig:unbuffered-comparison} shows
the slowdown of all data types compared to the algorithm on \texttt{float}.
As shown earlier, the slowdown of \repro{ScalarT}{L} ranges
from a factor%
\footnote{For the omitted configuration of $\text{\texttt{L}}=4$,
          the slowdown is even up to \factor{12}.}
\numrange{4}{10} for small numbers of groups
compared to built-in \texttt{float}s
(which has the same performance as \texttt{DECIMAL(9)})
and steadily decreases to a factor \numrange{1.5}{3}
as the numbers of groups increases.
The latter effect is due to the fact that the total runtime increases
by the same constant amount for all data types
(the partitioning is exactly the same for all of them)
with an increasing number of groups,
while the overhead of \texttt{operator+=}
is more or less independent of the number of groups,
making the former relatively more dominant than the latter.

\textbf{Conclusion:}
Using reproducible floats as drop-in replacement
in unmodified state-of-the-art aggregation algorithms
has an overhead of up to factor 6
compared to built-in floats with comparable accuracy.

\subsection{Aggregation with Summation Buffers}
\label{sec:eval-buffered-hashagg}

We now turn our attention to \algo{PartitionAndAggregate}
\emph{with} summation buffers.
We first show micro-benchmarks
that confirm our choice of the buffer size $\varname{bsz}$
and then evaluate end-to-end performance.

In order to confirm our model of the cache footprint
that leads to Equation~\ref{eqn:2pass-bsz},
we compare the performance of \algo{PartitionAndAggregate}
for several values of $\varname{bsz}$
and $d=0$, i.e., no partitioning.
Figure~\ref{fig:buffersize-pass1} shows the results.
According to Figure~\ref{fig:buffersize-pass1}a and
like for the isolated summation routine in Section~\ref{sec:tuning-summation},
for a very small number of groups,
the larger the buffer is, the better is the performance.
However, performance gains of buffers larger than $2^8$ elements are only marginal.
As Figure~\ref{fig:buffersize-pass1}b shows,
the situation is different for slightly larger numbers of groups.
Here performance sharply drops for buffers larger than $2^8$ and $2^7$ elements
for single-precision and double-precision data types respectively.

This is in line with our model of the working set
introduced in Section~\ref{sec:parameter-tuning}:
For all configurations shown in
Figure~\ref{fig:buffersize-pass1}b,
the performance drops when the modeled working set exceeds \SI{1}{\mebi\byte},
which is about half of the last-level cache \emph{per core}.
Figure~\ref{fig:buffersize-pass1}c details for which number of groups
the caching effect starts to be noticeable for a given buffer size
using the \repro{float}{2} data type:
For each fixed buffer size,
the performance drops sharply once a certain number of groups is reached.
Again, this is the point when the working set
exceeds \SI{1}{\mebi\byte} according to the equation
of Section~\ref{sec:parameter-tuning}.
The performance of the other data types follows the same pattern.

If we compare the value of $\varname{bsz}$
predicted by Equation~\ref{eqn:2pass-bsz}
with the buffer size performing best in Figure~\ref{fig:buffersize-pass1}c,
we can see that the predicted value
is very close to the optimal in all situations,
with only a few exceptions
(for example, $\varname{bsz}=512$ is slightly better
than the predicted $\varname{bsz}=1024$ for $2^6$ groups).
However, we found through exhaustive search that \SI{75}{\percent}
of all combinations of number of groups and data types
deviate by less than \SI{1}{\percent} from the optimal performance,
\SI{90}{\percent} of them deviate by less than \SI{5}{\percent},
and the largest deviation is \SI{20}{\percent}.
Appendix~\ref{app:buffersize-pass2} shows the same procedure
for \algo{PartitionAndAggregate} with $d=1$,
where our model predicts the optimal buffer size equally well
(the percentiles of the slowdown are even slightly lower).

\textbf{Conclusion:}
We can predict a close-to-optimal buffer size
using Equation~\ref{eqn:2pass-bsz},
which we use for the remainder of this paper.

\smallskip

\begin{figure}
    \includegraphics{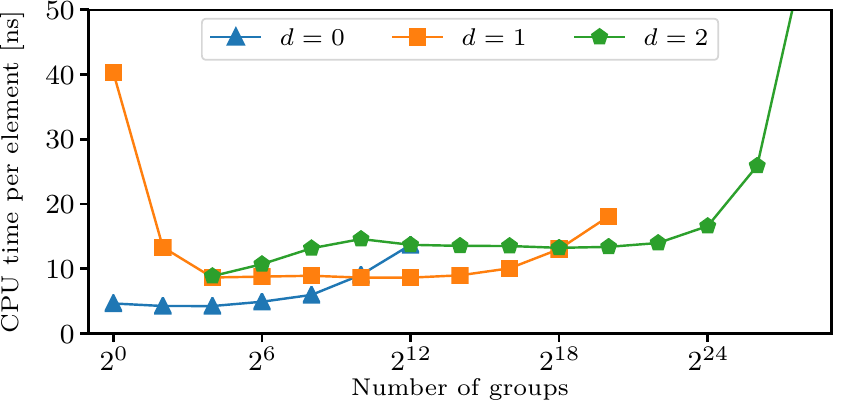}
    \caption{\algo{HashAggregation} variants with different amount of
        partitioning on \repro{float}{2}.}
    \label{fig:tune-passes}
\end{figure}

In order to determine how much partitioning is needed as pre-processing,
we compare \algo{PartitionAndAggregate}
for different number of levels of partitioning.
Each variant uses the optimal value of $\varname{bsz}$
as predicted by Equation~\ref{eqn:2pass-bsz} for the given number of groups.
Figure~\ref{fig:tune-passes} shows the result for \repro{float}{2}:
By comparing the performance of the three variants,
we can see that each level of partitioning has additional costs,
which is only worth it if the number of groups is so large
that the inefficiency due to smaller buffers (or out-of-cache processing)
is even more expensive.
Concretely, no partitioning at all is faster
as long as the number of groups is less than $2^{10}$.
After that point, partitioning once is faster,
so the partitioning pays off.%
\footnote{If the number of partitions is smaller than the number of threads,
          some threads are idle while the others aggregate their partitions,
          which explains the performance drop in these cases.
          However, the final algorithm uses \algo{PartitionAndAggregate}
          only for much larger numbers of groups
          and is, hence, not affected by this phenomenon.}
Similarly, two levels of partitioning are faster than just one
for $2^{18}$ groups or more.
This corresponds to $2^{10}$ groups per partition---%
so the two thresholds are effectively the same.

More levels of partitioning are not beneficial
for the data sets with $2^{30}$ records used in this paper,
but may be helpful for even larger ones.
The performance drop with more than $2^{24}$ is not due
to the number of groups becoming too large,
but due to the number of records per group becoming too small.
We discuss this effect in more detail
in Appendix~\ref{app:perf-distinct-data}.

In experiments not shown in this paper,
we determine the thresholds on other data types as well.
The results are largely the same, so we omit them here.
Furthermore, we only do the tuning
with numbers of groups that are powers of two.
We could determine the thresholds more precisely,
but since small changes in their values
do not have a drastic impact on performance,
we do not implement that in our prototype.

\textbf{Conclusion:}
Partitioning the input helps to reduce the cache footprint of our algorithm,
which in turn improves performance.
The more groups there are in the input,
the more levels of partitioning are needed
and our micro-benchmarks show the cross-over points.

\smallskip

\begin{figure}
    \includegraphics[width=\columnwidth]{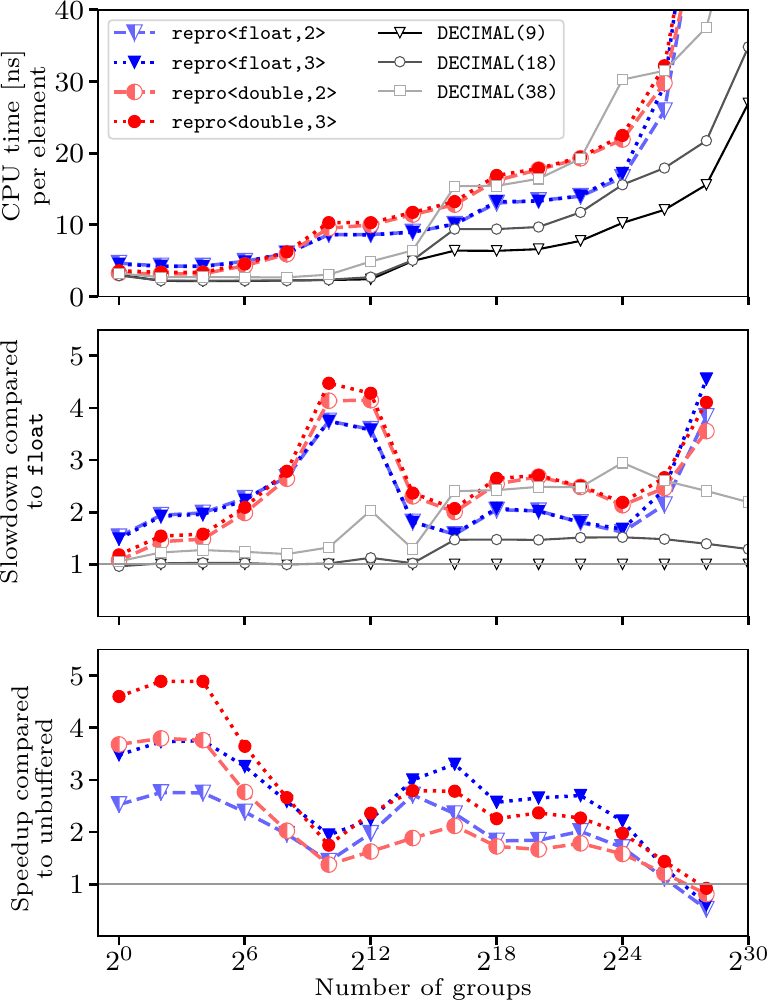}
    \caption{\algo{PartitionAndAggregate}
             with summation buffers on various \repro{ScalarT}{L}
             compared to the same algorithm on unbuffered \texttt{DECIMAL}.}
    \label{fig:final-comparison}
\end{figure}

Figure~\ref{fig:final-comparison} shows
the performance of \algo{PartitionAndAggregate}
using \repro{ScalarT}{L} and summation buffers
in comparison with unbuffered \texttt{DECIMAL} types.
The upper diagram shows the absolute running time,
which exhibits the same pattern of increasing cost
due to more levels of partitioning for increasing numbers of groups.
Compared to the algorithm without summation buffers, however,
the running time is generally lower and, in particular,
there is now little difference
between different configurations of \repro{ScalarT}{L}.
The largest difference is caused by the fact
that the reproducible data types based on \texttt{double}
are slower than those based on \texttt{float}.
This is mainly due to the fact
that \algo{Partitioning}, which is memory bound,
needs to move twice as much data in the former case.
This is the same effect that slows down the \texttt{DECIMAL} types,
which makes them about as slow or slower as our reproducible types
for $2^{16}$ groups and more
(in addition to being less flexible).
Finally, starting at around $2^{24}$ groups,
where the keys become more and more distinct,
performance starts dropping considerably,
an effect that we explain in more detail
in Appendix~\ref{app:perf-distinct-data}.

\itodo{ingo: This is true, but also, my implementation is inefficient
       with another level of partitioning.
       I hope we can get away with only three levels\ldots}
If we compare this result with the known results
for algorithms on built-in types~\cite{Cieslewicz2007,Muller2015},
we see that the algorithms here run out of cache
already for much smaller numbers of groups.
This is due to the larger cache footprint
of the summation buffers as argued earlier.
Running out of cache has also a stronger effect
caused by the additional indirection
(accessing the offset and the end of the buffer
may cause two cache misses instead of just one).

The middle diagram shows what the absolute performance means
in terms of slowdown compared to built-in \texttt{float}s.
We can see that the summation buffers
have significantly narrowed the gap between our new reproducible data types
and the built-in floating"-point types.
In many cases, this slowdown is in the range of \numrange{1.3}{2.5}.
Only for almost distinct data (more than $2^{26}$ groups)
and for the range of $2^8$ to $2^{12}$ groups,
the slowdown may be larger than that.
In the latter range, the number of groups are such that
the algorithm on built-in floats can still fit its working set into the last-level cache,
while the algorithm using summation buffers cannot,
and thus needs to pay the additional price of partitioning the input first
(on top of the overhead of buffering and more expensive summation).

\begin{table}
  \centering
  \begin{filecontents*}{data/slowdown-gmean-MultiPhaseHashIncache.csv}
data_type;slowdown
repro<double,1>;2.120778673721867
repro<double,2>;2.1845718792166804
repro<double,3>;2.2948959401045026
repro<double,4>;2.4089830740132987
repro<float,1>;1.8796214525734827
repro<float,2>;2.108566805752429
repro<float,3>;2.161296747671356
repro<float,4>;2.35091920535119
\end{filecontents*}
\pgfplotstabletypeset[
      texttt/.style={%
        preproc cell content/.append style={/pgfplots/table/@cell content/.add={\tt}{}},
      },
      col sep=semicolon,
      every head row/.style={%
        before row={\toprule},
        after row=\midrule,
      },
      every last row/.style={after row=\bottomrule},
      columns/data_type/.style={string type,column type=l,texttt,column name=data type},
    ]{data/slowdown-gmean-MultiPhaseHashIncache.csv}
  \caption{Geometric mean of slowdown of summation buffers
           compared to unbuffered per data type.}
  \label{tbl:final-slowdown-gmeans}
\end{table}

Table~\ref{tbl:final-slowdown-gmeans} shows
that the slowdown is still reasonable:
its geometric mean of all numbers of groups
ranges from \numrange{1.87}{2.35} for types based on \texttt{float}
and from \numrange{2.12}{2.41} for types based on \texttt{double}.
We believe that this is an affordable price for full reproducibility.

Finally, the lower diagram shows the speedup
of our algorithm with summation buffers
compared to the naïve approach without them.
In particular for small data sets,
the speedup is considerable
(between factor 2 and more than 5 for the shown configurations
and up to factor 6 for the omitted $\text{\texttt{L}}=4$).
As expected, it is the higher, the larger \texttt{L}.
The speedup drops slightly below 1 for the largest number of groups,
i.e., using summation buffers is actually slower than not using them.
Since the difference is not large, we leave this as a small open problem.

\textbf{Conclusion:}
Thanks to efficient partitioning routines, careful cache-management,
and vectorized summation on summation buffers,
the overhead of reproducibility on floating-point numbers
can be reduced to a slowdown of about a factor of two.

\itodo{Ingo: Should I try out an GMP implementation?}
\itodo{Stefan: You should have a non-reproducible test case.}

\vspace{-2mm}
\subsection{End-to-End Query Performance}

We integrated our reproducible data types
into MonetDB~\cite{Boncz2008} v11.25.23
in order to quantify their impact on end-to-end query performance.
To that aim, we modified MonetDB's aggregation operator
for sum on built-in doubles
such that it first aggregates its input into a locally allocated array
using our reproducible data types (with or without summation buffers)
and then copies the result converted to doubles
into the result array allocated by the system.%
\footnote{This does not technically make MonetDB reproducible
          because it parallelizes query plans
          as independent subplans on parts of the input
          whose intermediate results are merged.
          Our changes make the aggregation operators
          of each subplan as well as the merging reproducible,
          but the splitting of the input remains non-deterministic.
          We argue that this still gives
          a good approximation of the performance impact.
          A full integration would require the introduction of a new type,
          which is a development effort out of the scope of this paper.}
We run a modified TPC-H benchmark as workload
where we replaced all \texttt{DECIMAL} columns by \texttt{DOUBLE}.

\begin{table}
  \centering
  \begin{tabular}{@{}lSSSS@{\hspace{-1.6em}}}
    \toprule
            & \textbf{\texttt{double}}
            & {\lbcell{\textbf{\repro{d}{4}} \\ without buffer}}
            & {\lbcell{\textbf{\repro{d}{4}} \\ with buffer}}
            & {\lbcell{\textbf{\texttt{double}} \\ (sorted)}}\hspace{1.6em} \\
    \midrule
      Aggregations & 34.2 & 51.3 & 38.7 & 45.1\\
      Other        & 65.8 & 63.1 & 64.0 &682.1 \\
    \midrule
      Total        &100.0 &114.4 &102.7 &727.2 \\
    \bottomrule
  \end{tabular}
  \caption{CPU time of different approaches for TPC-H Query~1
           relative to the total CPU time on built-in doubles in \%.}
  \label{tbl:tpch-q1}
  \vspace{-1mm}
\end{table}


Table~\ref{tbl:tpch-q1} shows the CPU time of different approaches on Query~1
relative to the CPU time of an unmodified MonetDB.
As additional baseline, we include the CPU time of modified queries
that sort the input to the grouping and aggregation operators,
which is the only way to make them reproducible across input permutations
without modifying the system.

As the table shows, using \repro{double}{4} without summation buffers
takes about \SI{14}{\percent} longer
due to a \SI{50}{\percent} CPU time increase of the aggregation operators.
This increase is lower than the \SI{10}{\x} increase
observed with our own aggregation operator
in Section~\ref{sec:eval-naive-integration}
due to the slower baseline of MonetDB's operator,
which performs several overflow checks for each input element.
With summation buffers, however, the overhead of reproducibility
is a negligible \SI{2.7}{\percent}.
Sorting, in contrast, is more than \SI{7}{\x} slower,
which shows the importance of a numeric solution such as the one we propose.

\section{Related Work} 


Aggregation with \algo{GroupBy} on conventional data types
is a well understood problem.
In recent years, it has been studied extensively
for in-memory, multi-core database systems%
~\cite{Cieslewicz2007,Ye2011,Muller2015,Wen2013a}.
The focus of that work was
contention-free parallelization and cache efficiency.
Which strategy is best to achieve the latter goal
mainly depends on the size of the result,
which directly depends on the number of groups.
The consensus~\cite{Cieslewicz2007,Muller2015} is that algorithms
similar to \algo{PartitionAndAggregate} with various levels of partitioning
are best for different numbers of groups,
which is why we build on them in this paper.
For the case where the result is larger than a private cache,
but smaller than the combined shared cache of all threads,
\textcite{Cieslewicz2007} show that
\algo{SharedAggregation} may be a better solution than the other two,
which uses uses a shared (lock-free) hash table,
at least in the absence of skew.
Similar techniques have been proposed
for \algo{Join} and \algo{Sort} operators%
~\cite{Balkesen2014,Balkesen2013,Balkesen2013a,Boncz2008,Barthels2017}.
As we show in this paper,
these techniques alone are not sufficient
for reproducible floating-point numbers.

Since the number of groups is generally not known in advance
and hard to estimate,
researchers have proposed mechanisms
to select the processing strategy adaptively~\cite{Muller2015,Cieslewicz2007},
which is possible with minimal overhead.
Furthermore, mechanisms for handling data skew in the input
or a constrained amount of memory
were proposed~\cite{Muller2016,Muller2015,Cieslewicz2007}.
These aspects are somewhat independent of reproducibility
and the proposed solutions can be applied to our algorithms as well,
so we do not go into further detail in this paper.

Variants of \algo{SortAggregation},
have not been found competitive by recent studies~\cite{Balkesen2014}
(except for presorted inputs~\cite{Cieslewicz2007}),
mainly due to much higher computational costs
and the difficulty to combine early aggregation with vectorization.

There is a line of research in High-Performance Computing
that studies the problem of reproducibility of numerical computations
using floating-point numbers%
~\cite{Demmel2013,Arteaga2014,Demmel2015}.
However, as argued throughout the paper,
the proposed solutions are not applicable to aggregation with \algo{GroupBy}.
No work on numeric reproducibility in the field of data processing
is known to us.

Our work may seem to contradict attempts to speed up query execution
either by approximating the computation or reducing the precision.
Prominent examples of the first class include
DBO~\cite{Jermaine2008}, BlinkDB~\cite{Agarwal2013},
and Sample+Seek~\cite{Ding2016},
as well as recent summarization techniques
based on the principle of Maximum Entropy~\cite{Orr2017}.
We consider this work somewhat orthogonal to ours
since many sampling techniques are based
on deterministic pseudo-random number generators
and could require a technique similar to ours to be completely reproducible.
Maybe more importantly, from a user's perspective,
it is clearly less surprising
to get different results from a system that gives approximate answers by design
rather than from one that is assumed to be deterministic.
Examples of the second class include
work on neural networks with 8-bit and even 1-bit precision
by \textcite{Suda2016} and \textcite{Courbariaux2016}, respectively,
as well as the low-precision machine learning framework ZipML~\cite{Zhang2016}.
However, as discussed in Section~\ref{sec:non-solutions},
precision and reproducibility are completely orthogonal
(our algorithms could be implemented
based on lower-precision floating-point types as well).

\section{Summary and Conclusion} 
\label{sec:conclusion}

In this paper we have addressed the problem
of bit"-reproducible aggregation in database systems.
The main challenge is that achieving reproducibility is expensive
and cannot be efficiently done with existing algorithms
for reproducible summation and for \algo{GroupBy}.
Any naïve combination of existing results in the two areas
leads to prohibitive overheads. 

The main insights from the work include identifying the bottlenecks
that result from the bookkeeping needed to keep track of rounding errors
and the effects that it has in cache locality for high cardinality aggregation.
Based on these insights,
we have proposed ways to extend existing aggregation operators
with bit"-reproducibility in a way
that the resulting overhead is acceptable
and comparable to that of conventional aggregation over built-in types.

With these results, we establish the basis
for exploring more complex data types and operations inside the database engine
providing the same guarantees and meeting the same requirements
as those imposed on regular code processing floating-point data.
As part of future work we intend to look into operators for machine learning,
vector manipulation, and series analysis
based on the algorithms presented in this paper. 

\section*{Acknowledgments}

We thank Eric Sedlar (Oracle Labs)
for bringing the problem of reproducibility in the context of database systems
to our attention, as well as for valuable feedback on this work.
We also thank Lefteris Sidirourgos for his feedback
on the integration of our algorithm into MonetDB.

\printbibliography 

\appendix

\section{Additional Experiments} 

In this appendix, we show two more experiments
that allow interested readers to understand our implementation in more depth.

\subsection{Performance of Partition and Aggregate on Almost Distinct Data}
\label{app:perf-distinct-data}

\begin{figure}[b]
    \includegraphics[width=\columnwidth]{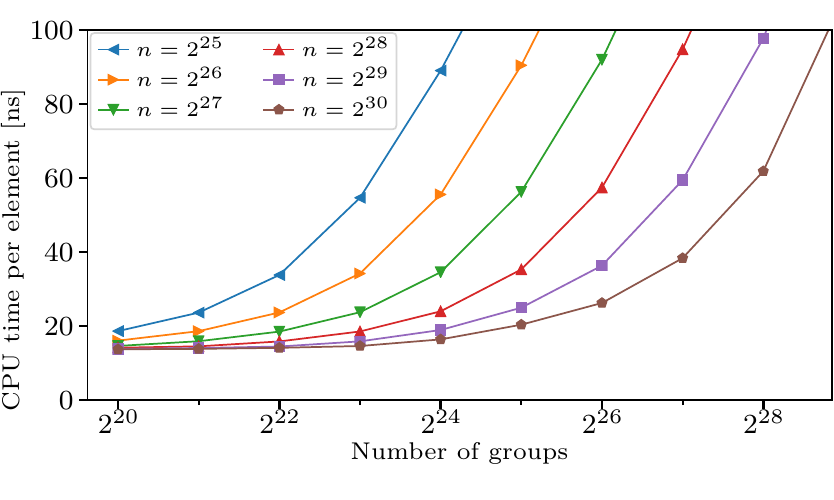}
    \caption{\algo{PartitionAndAggregate} with $\varname{bsz}=256$
    for various input sizes on \repro{float}{2}.}
    \label{fig:distinct-data-problem}
\end{figure}

\begin{figure*}
    \centering
    \includegraphics[width=\textwidth]{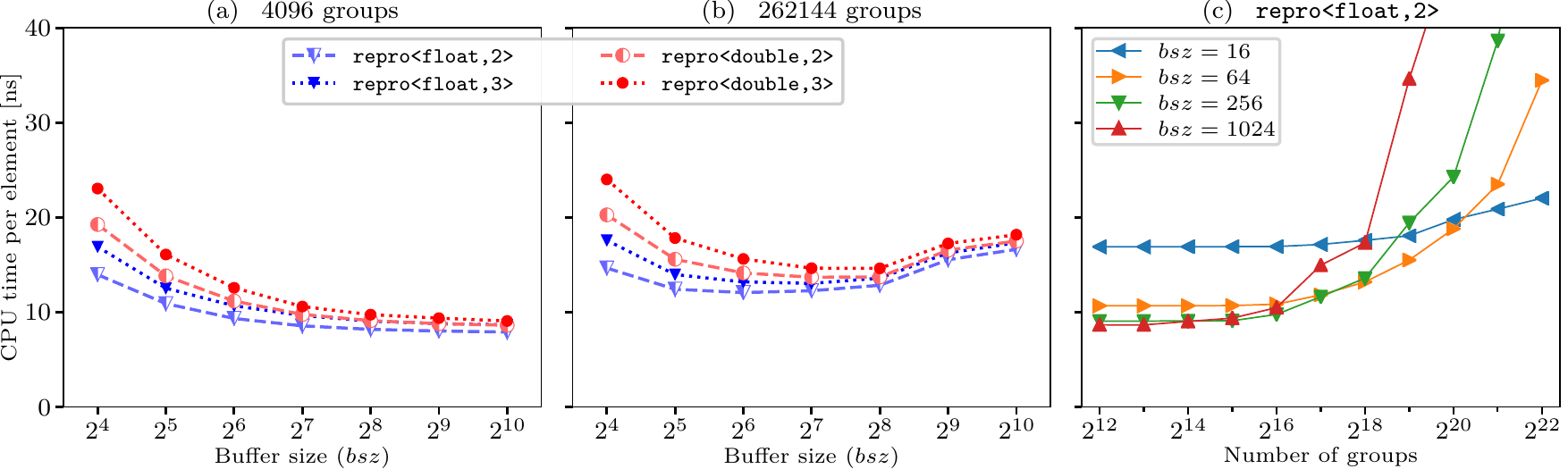}
    \caption{\label{fig:buffersize-pass2}
             Impact of the buffer size on \algo{PartitionAndAggregate}
             with a partitioning fanout of 256 and $d=1$.}
\end{figure*}
\setlength{\textheight}{12cm}

Figure~\ref{fig:final-comparison} in Section~\ref{sec:eval-buffered-hashagg}
shows a performance drop of \algo{PartitionAndAggregate}
for almost distinct data,
i.e., when the number of groups approaches the number of records in the input.
Here, we explain the reason for this drop and sketch a possible solution.
Figure~\ref{fig:distinct-data-problem} shows the performance of our algorithm
for distinct inputs of various sizes.
As the plot shows, the performance drops
whenever the average number of records per group
$n/n_{\varname{groups}} < 2^6$ independently of the input size.
Several effects add up in these situations:
First, the summation routine is less efficient for fewer elements per call
as discussed in Section~\ref{sec:tuning-summation}.
Second, as with the algorithm on IEEE floats,
the cost of bringing the memory for the final result into cache
and writing it back to RAM
becomes noticeable as the result size increases.
And last, the fact that our reproducible algorithm
first produces local aggregates using summation buffers
and only then writes them to the result
represents additional costs that increase linearly with the number of groups
and become dominant after some point.
We believe that our algorithm can be improved upon for these cases,
but leave such an optimization for future work.

\itodo{ingo: Andrea and I tried hard to fix this problem,
       but we are not sure, what exactly the problem is.
       Tuning to small buffer sizes as in Section~\ref{sec:tuning-summation}
       did not help.
       What we could do is to use the unbuffered algorithm for these cases.
       \algo{SortAggregation} is another option we could try out.
       There are also some implementation tricks that I could still try out.}

\subsection{Tuning the Buffer Size in Partition and Aggregate}
\label{app:buffersize-pass2}

Figure~\ref{fig:buffersize-pass2} shows the impact
of the buffer size ($\varname{bsz}$)
on the running time of \algo{PartitionAndAggregate}
for various numbers of groups in the input.
Qualitatively, the impact is exactly the same as without prior partitioning
(see Figure~\ref{fig:buffersize-pass1}).
However, the partitioning divides the number of groups
processed at the same time by the partitioning fan-out (256),
i.e., data sets with 256 times more groups
can be aggregated before the working set exceeds the cache.
At the same time, the running time is increased
by the constant costs of the partitioning routine,
which is independent of the buffer size or the number of groups.

\end{document}